\documentclass{aa} 
\usepackage{graphicx, color}
\usepackage{txfonts}
\usepackage{orcidlink}
\begin{document} 

\title{The space weather around the exoplanet GJ 436\,b. \\ II. Stellar wind--exoplanet interactions}
   \titlerunning{The exo-space weather of GJ 436\,b. II}
   
   \author{A. A. Vidotto \inst{1}\orcidlink{0000-0001-5371-2675}
          \and   V. Bourrier \inst{2}\orcidlink{0000-0002-9148-034X}
          \and   R. Fares \inst{3}\orcidlink{0000-0002-3301-341X}
          \and S. Bellotti \inst{1,4,5}\orcidlink{0000-0002-2558-6920}
          \and J.~F. Donati \inst{5}\orcidlink{0000-0001-5541-2887}   
          \and P. Petit \inst{5}\orcidlink{0000-0001-7624-9222}  
          \and G.~A.~J. Hussain \inst{4}\orcidlink{0000-0003-3547-3783}
          \and  J. Morin \inst{6}\orcidlink{0000-0002-4996-6901} }

   \authorrunning{Vidotto et al.}
    
   \institute{Leiden Observatory, Leiden University, PO Box 9513, 2300 RA, Leiden, the Netherlands\\ \email{vidotto@strw.leidenuniv.nl}
	\and Observatoire Astronomique de l’Universit\'e de Gen\'eve, Chemin Pegasi 51b, 1290 Versoix, Switzerland
	 \and  Department of Physics, College of Science, United Arab Emirates University, P.O. Box No. 15551, Al Ain, UAE\\            \email{rim.fares@uaeu.ac.ae}
	 \and Science Division, Directorate of Science,  European Space Research and Technology Centre (ESA/ESTEC), Keplerlaan 1, 2201 AZ, Noordwijk, The Netherlands
	 \and Institut de Recherche en Astrophysique et Plan\'etologie,  Universit\'e de Toulouse, CNRS, IRAP/UMR 5277, 14 avenue Edouard Belin, F-31400, Toulouse, France 
	\and Laboratoire Univers et Particules de Montpellier, Universit\'e de Montpellier, CNRS,   F-34095, Montpellier, France}
   \date{Received ; accepted }
 
  \abstract
  {The M dwarf star GJ\,436 hosts a warm-Neptune that is losing substantial amount of atmosphere, which is then shaped by the interactions with the  wind of the host star.  The stellar wind is formed by particles and  magnetic fields that shape the exo-space weather around the exoplanet GJ\,436\,b. Here, we use the recently published magnetic map of GJ\,436 to model its 3D Alfv\'en-wave driven wind. By comparing our results with previous transmission spectroscopic models and measurements of non-thermal velocities at the transition region of GJ\,436, our models indicate that the wind of GJ\,436 is powered by a smaller flux of Alfv\'en waves than that powering the wind of the Sun. This suggests that the canonical flux of Alfv\'en waves assumed in solar wind models might not be applicable to the winds of old M dwarf stars. Compared to the solar wind, GJ\,436's wind has a weaker acceleration and  an extended sub-Alfv\'enic region. This is important because it places the orbit of GJ\,436\,b inside the region dominated by the stellar magnetic field (i.e., inside the Alfv\'en surface). Due to the sub-Alfv\'enic motion of the planet through the stellar wind, magnetohydrodynamic waves and particles released in reconnection events can travel along the magnetic field lines towards the star, which could power the anomalous ultraviolet flare distribution recently observed in the system. For an assumed planetary magnetic field of $B_p \simeq 2$~G, we derive the power released by stellar wind-planet interactions as $\mathcal{P} \sim 10^{22}$ -- $10^{23}$~erg~s$^{-1}$, which is consistent with the upper limit of $10^{26}$~erg~s$^{-1}$ derived from ultraviolet lines. We further highlight that, because star-planet interactions depend on stellar wind properties, observations that probe these interactions and the magnetic map used in 3D stellar wind simulations should be contemporaneous for deriving realistic results.
  }
  

   \maketitle

%%%%%%%%%%%%%%%%%%%%%%%%%%%%%%%%%%
\section{Introduction}
The evolution of planets is shaped by billions of years of interactions with their host stars and also with neighbouring planets. In addition to the central gravitational force of the host star, there are several  types of interactions taking place between a star and an exoplanet \citep{2020IAUS..354..259V}, some of which are not present, or simply very weak, in the present-day solar system planets. 

 In the case of the GJ\,436 system, the focus of the current study, the warm Neptune GJ\,436\,b experiences a strong interaction with its host star due to the combined effects of high-energy stellar irradiation that is deposited in the planetary atmosphere, and subsequent interaction of this atmosphere with the stellar wind  \citep{2016A&A...591A.121B}. Spectroscopic transit observations in Lyman-$\alpha$ {\citep{2014ApJ...786..132K, 2015Natur.522..459E, 2017A&A...605L...7L, 2019A&A...629A..47D}} revealed that GJ\,436\,b is enshrouded by a giant cloud of escaping atmosphere that trails GJ\,436\,b along its orbit. Three-dimensional models suggest that this trailing structure takes the form of a giant comet-like tail \citep{2016A&A...591A.121B, 2018MNRAS.481.5315S, 2019ApJ...885...67K, 2021MNRAS.501.4383V}. By modelling the observations of spectroscopic transits of GJ\,436\,b with atmospheric escape models that account for atmospheric interaction with the stellar wind, local properties of the stellar wind at the orbital distance of the planet, such as the wind speed  and density, can be estimated; these can then be used to infer global properties of stellar winds, such as their mass-loss rates \citep{2017MNRAS.470.4026V}. All these quantities shape the space weather around a planet. 

Indeed, one of the major strengths in observing  and modelling star-planet interactions in general is that they can provide alternative ways to physically characterise planetary systems. Stellar wind mass-loss rates, which are rather difficult to constrain in the case of cool dwarf stars \citep{2021LRSP...18....3V}, can also be estimated by investigating star-planet interactions through planet-induced radio emission \citep{2021MNRAS.504.1511K, 2023NatAs.tmp...65P} or auroral radio emission from the exoplanet \citep{2019MNRAS.488..633V}. Investigating star-planet interactions through chromospheric hot spots associated to anomalous stellar activity \citep[e.g.][]{2005ApJ...622.1075S, 2019NatAs...3.1128C} or exoplanetary radio emission \citep[e.g.][]{1999JGR...10414025F, 2021A&A...645A..59T} allows one to probe planetary magnetism, a quantity that is believed to be very important in the context of planetary habitability, but is still rather poorly known in planets outside the solar system. 

In this work, we focus on star-planet interactions that are mediated by the stellar wind and its embedded magnetic field. In the literature, this type of interaction can be found as `star-planet (electro)magnetic interactions',  `stellar wind--planet interactions' or `electrodynamic star-planet interactions', but they are only different names of the same process. In the context of star-planet interactions mediated by the magnetised host star's wind, different physical processes take place depending on the wind regime in which the planet orbits.

In the regime of sub-Alfv\'enic motion of the planet through the wind of its host star, a magnetic `{connectivity}' between the planet and the star can take place. In this regime, the orbiting planet can trigger magnetohydrodynamics (MHD) waves that can travel towards the star, or alternatively (and even possibly concurrently) it can also trigger magnetic reconnection events in case of a magnetised planet. Both of these mechanisms release energy that are radiated away and could be detected in the form of, e.g., chromospheric hot spots \citep{2009A&A...505..339L, 2022MNRAS.512.4556S},  planet-induced coronal radio emission \citep{2013A&A...552A.119S, 2023NatAs.tmp...65P}, anomalous flare events \citep{2019ApJ...872..113F}, and even potentially the lack of energetic flare events as recently proposed by \citet{2023AJ....165..146L} in the context of GJ\,436. Note that all these signatures originate on or near the host star, and not on the planet.

The inverse takes place in the regime of super-Alfv\'enic motion of the planet through the wind of its host star. In this regime, the interaction between a planet and the stellar wind can lead to radio emission originating from the planet itself (more precisely, from its magnetosphere). This is because in the super-Alfv\'enic regime, the magnetic {connectivity} between the star and planet can no longer form. This is the situation experienced by the solar-system planets currently, where we see that the planetary auroral radio emission is correlated with the solar wind energy that is dissipated in the magnetosphere of  solar-system planets \citep[e.g.,][]{1999JGR...10414025F}. In spite of many attempts to detect such signature in exoplanets, it was only recently that a potential detection of radio emission from a hot Jupiter was reported (\citealt{2021A&A...645A..59T}, see also \citealt{2023A&A...671A.133E} for models of these observations).

Fundamental to the nature of stellar-wind--planet interactions is thus the regime of the orbital motion of the planet, i.e., sub- or super-Alfv\'enic. To determine such a regime, we need to obtain more realistic stellar wind properties, such as densities, velocities and magnetism. These stellar wind quantities are also what shape the space weather of planets, and are thus relevant also for, e.g., understanding the flux of energetic particles impacting planetary atmospheres \citep[e.g.,][]{2021MNRAS.505.1817M, 2023MNRAS.521.5880R}. In the case of GJ\,436, its wind has been investigated in several works \citep{2017MNRAS.470.4026V, 2020MNRAS.494.1297M, 2021MNRAS.505.1817M}, but it is only recently that the surface magnetic field of the host star GJ\,436 was mapped \citep{2023arXiv230615391B}, which allows us to now model for the first time the three-dimensional (3D) structure of the stellar wind including the observed surface magnetism. Magnetic fields  play an important role in the winds of cool dwarf stars, as  they are responsible for the  mechanisms that heat and accelerate these winds. Section \ref{sec.model} presents our  3D MHD simulations of the wind of GJ\,436. We then use the results of our wind models to compute the space weather conditions (Section \ref{sec.spaceweather}) along the highly misaligned orbit of GJ\,436\,b. GJ\,436\,b is a Neptune-mass planet orbiting very close to its slowly rotating M dwarf host. It has a puzzling observed orbital architecture -- non-circular orbit with an eccentricity of $0.152$ \citep{2018A&A...609A.117T} and strong orbital misalignment \citep{2018Natur.553..477B, 2022A&A...663A.160B} -- that is likely the result of a delayed migration following Kozai-Lidov interactions with an outer massive companion  \citep{2018Natur.553..477B, 2021A&A...647A..40A}. In Section \ref{sec.spi}, we then compute the  strengths of star-planet interactions. In the final section of this paper (Section \ref{sec.conclusion}), we discuss how our computations compare to the recent observations of \citet{2023AJ....165..146L} and present our concluding thoughts.

%%%%%%%%%%%%%%%%%%%%%%%%%%%%%%%%%%%%%%
\section{The wind of GJ\,436}\label{sec.model}
\begin{figure}
    \centering
    \includegraphics[width=.48\textwidth]{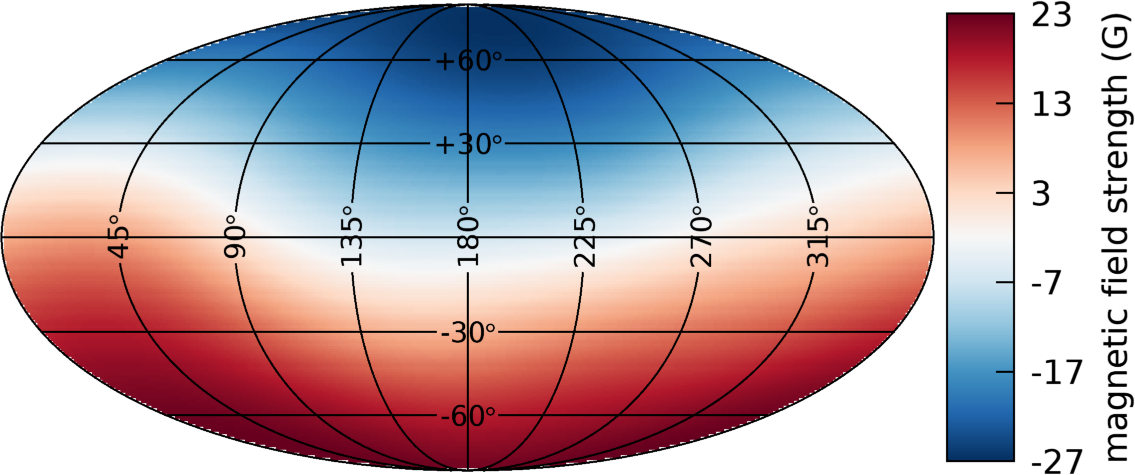} 
    \caption{Observationally reconstructed radial magnetic field of GJ\,436 from \citet{2023arXiv230615391B} {using Zeeman Doppler Imaging}. This map is incorporated at the boundary of our stellar wind simulations.}
    \label{fig.zdi}%
\end{figure}

The environment surrounding GJ\,436\,b consists of particles and magnetic fields that form the wind of the host star. Here, we characterise this environment by performing data-driven 3D MHD simulations of the stellar wind. Because winds of cool dwarf stars are ultimately driven by stellar magnetism, at the boundary of our model (which is located close to the stellar surface, at the transition region), we include the reconstructed surface field shown in Figure \ref{fig.zdi} \citep{2023arXiv230615391B}. {This map was reconstructed using the Zeeman-Doppler Imaging technique \citep{1997A&A...326.1135D, 2009ARA&A..47..333D} and, thus, represents the large-scale field of GJ\,436 in March-June 2016.} For this simulation, we use the Alfv\'en Wave Solar Model (AWSoM, \citealt{2014ApJ...782...81V}) implemented in the numerical code BATS-R-US \citep{1999JCoPh.154..284P,2012JCoPh.231..870T}. We refer the reader to \citet{2014ApJ...782...81V} for further details of the model. Below we describe the main input parameters adopted in the simulations of the wind of  GJ\,436.

In addition to the stellar surface magnetic map, other model inputs are the stellar properties (mass, radius and rotation period; c.f.~Table\ref{table_properties}), the density and temperature at the transition region, and the Alfv\'en wave properties (boundary wave flux and the correlation length of the Alfv\'enic wave turbulence). Of these parameters, the least constrained ones are the wave properties, as we discuss below.

The inner boundary of our model is at the transition region, and  values considered for solar wind models \citep{2014ApJ...782...81V, 2017ApJ...845...98O} for ion/electron density ($n_{\rm tr} = 5 \times 10^{11}~ \mathrm{cm}^{-3}$), temperature ($T_{\rm tr}=5\times 10^4$~K), and, consequently, gas density ($P_{\rm tr} \simeq 1.4 ~\mathrm{dyn~cm}^{-2}$), were adopted. Our assumed pressure is not too dissimilar to the value assumed in the chromospheric models of the M dwarf star GJ832 by \citet{2016ApJ...830..154F}, who used $P_{\rm tr} = 0.8~\mathrm{dyn~cm}^{-2}$. For a temperature of $5\times 10^4$~K, this pressure value translates to a density of $10^{11}~ \mathrm{cm}^{-3}$, which is also similar to density values found at this temperature from the chromosphere+wind models of \citet{2021ApJ...919...29S}. However, there are also models in the literature that  adopt lower pressure values of $P_{\rm tr} = 0.02$ -- $0.04~\mathrm{dyn~cm}^{-2}$ \citep{1997A&A...326..249M, 2019ApJ...886...77P}, which would imply transition region densities of  a few times $ 10^{9}~ \mathrm{cm}^{-3}$ at a temperature of $5\times 10^4$~K. Depending on which chromospheric model we use to guide our assumption of $n_{\rm tr}$, we could have transition region densities that vary from $\sim 10^{9}$ to $10^{11} \mathrm{cm}^{-3}$. \citet{2021MNRAS.500.3438O} explored the effects that changing this parameter had on the wind of a sub-giant star $\lambda$~And and they found that varying $n_{\rm tr}$ by a factor of 100 led to similar wind properties (compare their models  A2 and B0, or B1 and D0, from their Table 2). We therefore do not expect that assuming a different value for $n_{\rm tr}$ within the quoted range would change our wind solutions substantially. 

\begin{table*}
\caption{Parameter adopted in our models for GJ\,436 and GJ\,436\,b.}
\label{table_properties}
\centering
\begin{tabular}{llllll} 
\hline
Parameter (Star) & value & units & reference\\
\hline
Mass ($M_\star$) & $\simeq 0.45$&$M_\odot$& \citet{2022MNRAS.514...77M}\\
Radius ($R_\star$) &  $\simeq 0.42$  &$R_\odot$& \citet{2022MNRAS.514...77M} \\
Rotation period  ($P_{\rm rot}$) & $\simeq 44.1$ & days& \citet{2018Natur.553..477B} \\
Stellar magnetism & c.f.~Fig.~\ref{fig.zdi} &  & \citet{2023arXiv230615391B} \\
Age & $6^{+4}_{-5}$ & Gyr& \citet{2007ApJ...671L..65T}\\
\hline
Parameter (Planet) & value & units & reference\\
\hline
Mass  ($M_{\rm pl}$) &21.72 &$M_\oplus$&  \citet{2022MNRAS.514...77M}\\
Radius  ($R_{\rm pl}$) &3.85& $R_\oplus$& \citet{2022MNRAS.514...77M} \\
Orbital period ($P_{\rm orb}$) & $\simeq 2.644$ & days&  Lanotte et al.~(2014)\\
Semi major axis ($a_{\rm orb}$) & 14.56 & $R_\star$ &  \citet{2022MNRAS.514...77M} \\
...  & $\simeq 0.028$ & au &  ...\\
Spin-orbit angle ($\Psi$) & $103.2$ & degrees &  Bourrier et al.~(2022)\\
\hline
\end{tabular}
\end{table*}

As they propagate through magnetic field lines, Alfv\'en waves transfer their energy and momentum to the wind plasma, thus causing its heating and acceleration. In AWSoM, the wave dissipation  is modelled through a turbulent cascade process. Its correlation length is assumed to increase with flux tube radius, thus $L_\perp = L_{\perp,0} \sqrt{B_\star/B}$. Here, we assume the proportionality factor being the same as that used in the AWSoM solar wind models, namely $L_{\perp,0}\sqrt{B_\star} = 1.5 \times 10^5~\mathrm {m}\sqrt{\mathrm{T}} = 1.5\times 10^4~\mathrm{km}\sqrt{\mathrm{G}}$. Based on the parametric study of \citet{2021MNRAS.500.3438O}, we also do not expect that changing the value of this base quantity by a factor of $\sim 10$ will substantially affect the general properties of our wind solution (see also \citealt{2018ApJ...853..190S}, who investigated the effects of the correlation length on wind properties such as its maximum temperature, mass-loss rate, and velocity).

An important quantity of our model is the boundary value of the Alfv\'en wave flux
\begin{equation}
S_{A_,0} =\rho_{\rm tr} \langle v_{\perp,0}^2 \rangle v_{A,0} \, ,
\end{equation}
where $\langle v_{\perp,0}^2 \rangle^{1/2}$ is the average velocity perturbation of the Alfv\'en wave, $\rho_{\rm tr} = m_p n_{\rm tr}$ is the mass density at the transition region, and the Alfv\'en velocity is $v_{A,0} = B_\star /\sqrt{4\pi \rho_{\rm tr}}$, where the index `0' indicates value computed at the boundary of the simulation. Here, $B_\star$ is the local value (i.e., at each longitude and latitude) of the surface magnetic field as derived in the surface magnetic map (Figure \ref{fig.zdi}). The previous equation can be rewritten as
\begin{equation}\label{eq.deltav}
\frac{\langle v_{\perp,0}^2 \rangle^{1/2}}{65~ \textrm{km/s}} =  \left( \frac{5 \times 10^{11}}{n_{\rm tr} [\textrm{cm}^{-3}]} \right )^{1/4}
 \left( \frac{S_{A,0}/B_\star [\textrm{erg/(cm}^2\textrm{\,s\,G})]}{1.1\times 10^7}\right)^{1/2}   \, .
\end{equation}
Note that the ratio $S_{A,0} / B_\star$ is assumed to be uniform at the inner boundary, implying that the average velocity amplitude of the Alfv\'en wave and the energy density of the wave ($\rho_{\rm tr} \langle v_{\perp,0}^2 \rangle$) are also uniform at the inner boundary. {The choice of solar wave flux adopted in previous models of the solar wind with AWSoM \citep{2013ApJ...778..176O, 2017ApJ...845...98O, 2014ApJ...782...81V} was guided by Hinode observations of velocity fluctuations at the chromospheric region \citep{2007Sci...318.1574D}. Indeed, }if one assumes that the non-thermal motions of ions at the transition region are associated to Alfv\'en waves \citep[e.g.,][]{2017ApJ...845...98O}, then one way to  constrain the Alfv\'en wave properties in other stars is by measuring the non-thermal broadening of lines formed in the transition region \citep{2023ApJ...950..124B}. In this case, the non-thermal velocity $\xi$ is related to the average velocity perturbation of the Alfv\'en wave as $\xi^2 = \frac12 \langle v_{\perp,0}^2 \rangle$. 

The ratio $S_{A,0} / B_\star$ is the most important free parameter in our model and its value  affects the wind temperature, its mass-loss rate and size of the Alfv\'en surface, which in turn affects the angular momentum loss of the star \citep{2020A&A...635A.178B, 2021MNRAS.500.3438O, 2021MNRAS.504.1511K}. Given that, we run two wind models with two different values of  $S_{A,0} / B_\star$: $1.1 \times 10^{6}~\textrm{erg (cm}^{2}\textrm{\,s\,G)}^{-1}$  and $1.1 \times 10^{7}~\textrm{erg (cm}^{2}\textrm{\,s\,G)}^{-1}$. The latter  is the same value as used in AWSoM solar wind simulations \citep{2013ApJ...764...23S, 2014ApJ...782...81V}. From now on, we refer to these models as `I' and `II'. As we will show in Section \ref{sec.spaceweather},  our derived mass-loss rates and computed non-thermal velocities of Model I show better agreement with observations, indicating that the adopted lower Alfv\'en wave flux is more appropriate to describe the wind of GJ\,436.

For GJ\,436, which has a surface radial field reaching about 28~G, our choice of parameters translates into correlation lengths of  $L_{\perp,0} \simeq 2800~\mathrm{km}$,\footnote{{The scaling of  the correlation length  $L_{\perp,0} \sqrt{B_\star} $ has been related to the distance between magnetic flux tubes on the solar surface \citep{1986JGR....91.4111H}.  \citet{1986JGR....91.4111H} empirically estimated a value of $L_{\perp,0} \sqrt{B_\odot} = 7520~{\rm km}\sqrt{\rm G}$ for the Sun, or $L_{\perp,0} \sim 4300$~km for $\sim 3$-G field.  We do not know how $L_{\perp,0}$ would change for GJ\,436, but because of its smaller radius, we naively expect that the distance between magnetic flux tubes would be smaller than solar for GJ\,436, consistent to the value of $\sim 2800~\mathrm{km}$ we use. Based on the parametric study of \citet{2021MNRAS.500.3438O}, we  do not expect a variation in this parameter by a factor of $\sim 10$ would substantially affect the general properties of our wind solution.}}  and  surface Alfv\'en fluxes of up to $S_{A,0}\simeq [3 \times 10^7, 3 \times 10^8] ~\textrm{erg cm}^{-2}{\mathrm{s}}^{-1}$, for Models I and II, respectively. The surface amplitudes of the wave perturbation are 21~km~s$^{-1}$ (Model I) and 65~km~s$^{-1}$ (Model II), uniform at the lower boundary. The corresponding surface energy densities of the wave are 11 and 35 erg~cm$^{-3}$. With these parameters, the predicted non-thermal velocities at the transition region are $\xi \simeq 14$ and $45$ km~s$^{-1}$, for Models I and II, respectively. {The other remaining parameters of the model, namely the stochastic heating parameter, the collisionless heat conduction parameter and the collisionless heat flux parameter have the same values as in Table 1 of \citet{2014ApJ...782...81V}. The energy equations are solved in their non-conservative form (i.e., solving for the pressure instead of total energy).}

\begin{figure*}[t]
    \centering
%    \sidecaption
    \includegraphics[width=.45\textwidth]{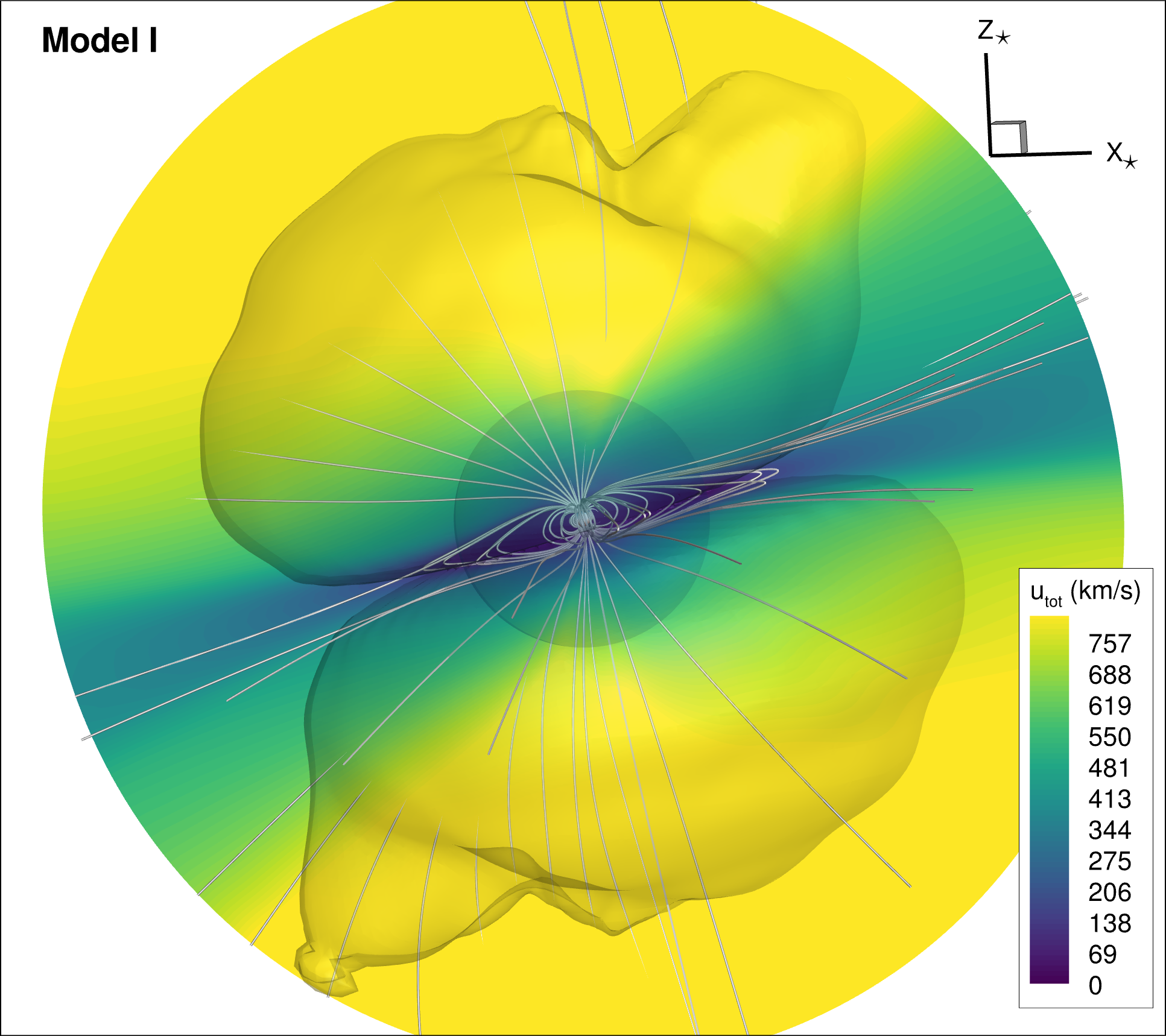}
      \includegraphics[width=.45\textwidth]{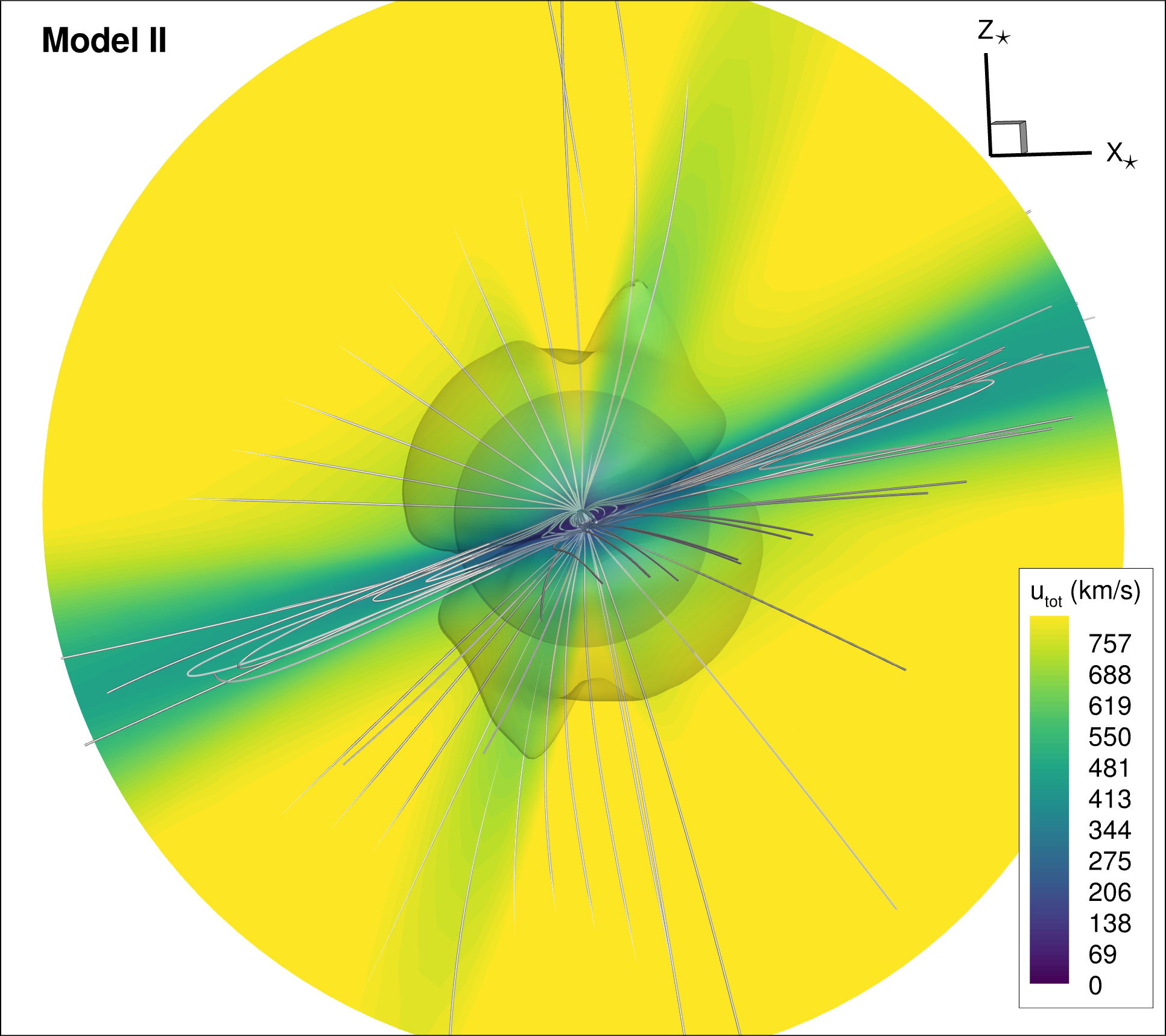}    \caption{Three dimensional view of the magnetised stellar wind for Models I and II. The surface magnetic field from March-June 2016 reconstructed by \citet{2023arXiv230615391B} is used at the boundary of the models. The grey streamlines are the magnetic field lines embedded in the wind.  The rotation axis of the star is along the $z_\star$-axis. The two-dimensional cut is placed at the $x_\star z_\star$ plane, and for indication, the orbital distance of the planet is shown by the translucent sphere at $14.56~R_\star \simeq 0.028$~au. The contour shows the wind velocity and the Alfv\'en surface is shown by the translucent outer surface. Models I and II predict stellar wind mass-loss rates of $1.1 \times 10^{-15}~M_\odot~\rm{yr}^{-1}$ and $2.5 \times 10^{-14}~M_\odot~\rm{yr}^{-1}$. }
    \label{fig:3d}%
\end{figure*}

Figure \ref{fig:3d} shows the 3D outputs of our simulations, after they have reached steady state. The grey streamlines represent the stellar wind magnetic field lines; this geometry results from the self-consistent interaction between field lines and the stellar wind flow. The rotation axis lies along positive $z_\star$ -- note that the magnetic field geometry is approximately that of an aligned dipole close to the stellar surface, with elongated closed field lines at large distances and open fields around the stellar rotation poles. The stellar wind mass-loss rate we obtained are $1.1 \times 10^{-15}~M_\odot~\rm{yr}^{-1}$ and $2.5 \times 10^{-14}~M_\odot~\rm{yr}^{-1}$, for Models I and II, respectively. For comparison, the solar wind mass-loss rate is $2 \times 10^{-14}~M_\odot~\rm{yr}^{-1}$ -- given that all of our input values for the Alfv\'en wave and transition region are solar for Model II, {and the surface magnetic flux of GJ\,436 is similar to solar,}\footnote{{From the magnetic map of GJ\,436, we find a surface magnetic flux of $\Phi_0 = 1.4 \times 10^{23}$~Mx, which is very similar to solar values (for an average large-scale field strength of $\sim 3$~G, the solar surface flux is  $\Phi_0 = 3 \times 4 \pi R_\odot^2 \simeq 1.8 \times 10^{23}$~Mx). In stellar wind theory, the mass-loss rate is linearly proportional to the open magnetic flux ($\Phi_{\rm open}$), which is also proportional to $\Phi_0$ \citep[e.g.,][]{2014MNRAS.438.1162V}. This is also seen in AWSoM models of winds of Sun-like stars \citep{2023MNRAS.524.2042E}. Because Model II uses the same solar wind parameters and GJ436 has a similar surface flux to the Sun, this model produces a mass-loss rate that is thus similar to that of the Sun.}} it is not surprising that the mass-loss rate derived in Model II is similar to that of the solar wind.
In Section \ref{sec.spaceweather}, we will perform a more detailed comparison of our derived mass-loss rates with those derived in other studies of GJ\,436. 
Figure \ref{fig:3d} also shows a cut in the $y_\star z_\star$ plane, where colours indicate the stellar wind speed, where we see that along open field lines the wind is accelerated more quickly, with low-speed streams around closed field line regions. As expected, higher velocities are achieved in Model II, with higher Alfv\'en wave fluxes. The Alfv\'en surface is shown by the translucent surface. For context, we plot a white circumference at the orbital distance of GJ\,436\,b (for simplicity, we assume a circular orbit). We note most of the time the planet would be in the region of open magnetic field lines and that the orbital distance is nearly always within the Alfv\'en surface for both models, which can have implications for star-planet interactions \citep{2020A&A...633A..48F} and induced radio emission on the star \citep{2021MNRAS.504.1511K, 2022MNRAS.514..675K}, for example. We will come back to this in Section \ref{sec.spi}. 

%%%%%%%%%%%%%%%%%%%%%%%%%%%%%%%%%%%%%%%%%%%%%%%%
\section{Space weather conditions of GJ\,436\,b}\label{sec.spaceweather}
\begin{figure}[!h]
    \centering 
    \includegraphics[width=.5\textwidth]{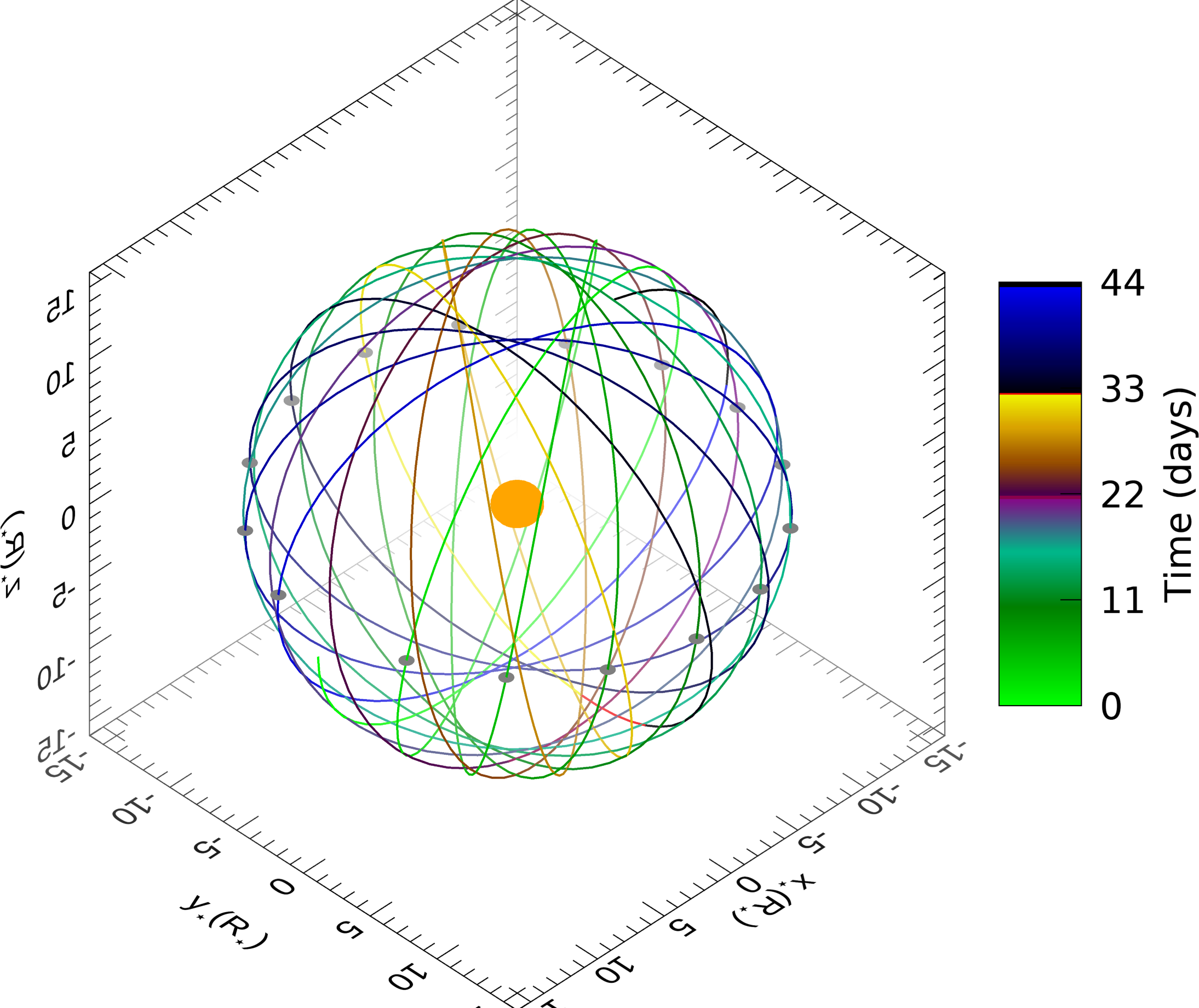}  
    \caption{Trajectories of the planet as seen in the reference frame of the star during one rotational period of the star (about 44~days, see colour bar), during which  the planet completes about 16 orbital revolutions. {The grey circles indicate the approximate times of mid-transit.}}
    \label{fig.pl_trajectory}%
\end{figure}

The space weather of GJ\,436\,b is formed by the particles and ambient magnetic field of the stellar outflow. Here, we compute the properties of stellar wind at the position of GJ\,436\,b. {
Because GJ\,436\,b has a very misaligned orbit,  we first compute the position of the planet as a function of time, following the coordinate transformations presented in Section \ref{sec.ref}.} To calculate the phase of the orbit with respect to the rotational phase of the star, we consider the transit midpoint at HJD$_t= 2458947.26212(12)$ \citep{2022MNRAS.514...77M}, with an orbital period of $  
2.64389803(27) $~d \citep{2014A&A...572A..73L}. 
At  stellar rotation phase zero (HJD$_s = 2457464.4967$, {corresponding to the first spectropolarimetric observation (\citealt{2023arXiv230615391B})}, and $\{ x_\star, y_\star\}=\{R_\star,0\}$ in our simulation coordinates), the planet was at an orbital phase of $\varphi_0=0.1745$ after the preceding transit midpoint. GJ\,436\,b has a nearly polar orbit with its orbital axis and stellar spin misaligned by 
$\Psi=103.2^\circ$ \citep{2022A&A...663A.160B}. Figure \ref{fig.pl_trajectory} shows the trajectories of the planet as seen in the reference frame of the star during one rotational period of the star (about 44~days), during which  the planet completes about 16 orbital revolutions.

\begin{figure*}
    \centering
    \includegraphics[width=.48\textwidth]{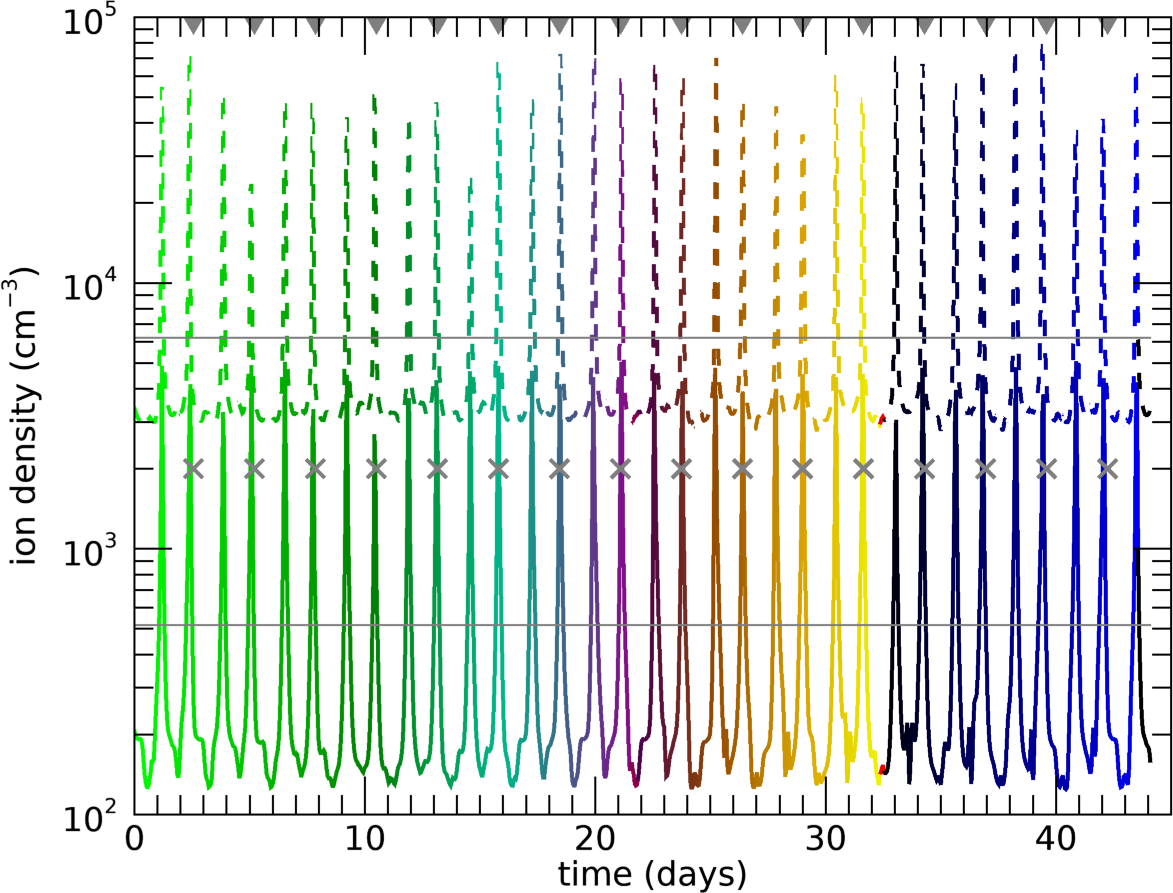}  
    \includegraphics[width=.48\textwidth]{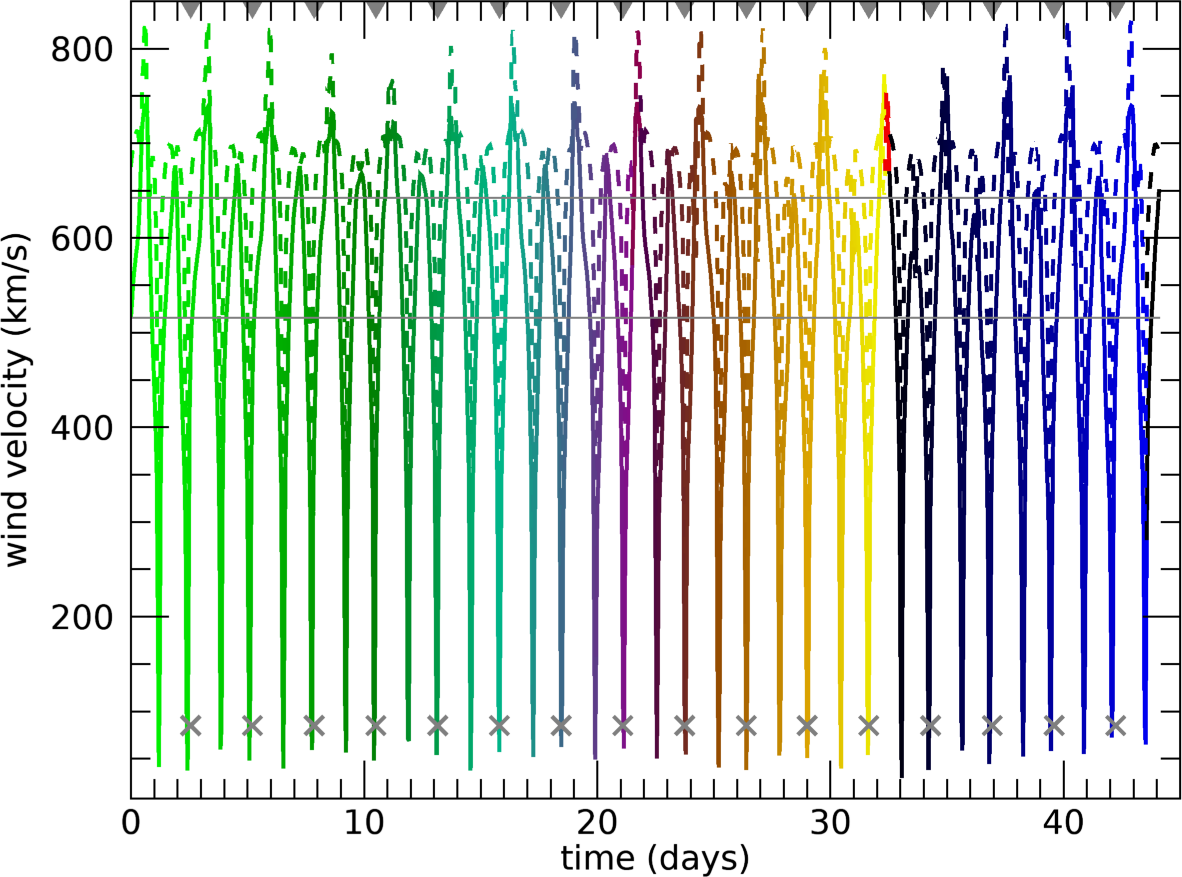} \\
    \includegraphics[width=.48\textwidth]{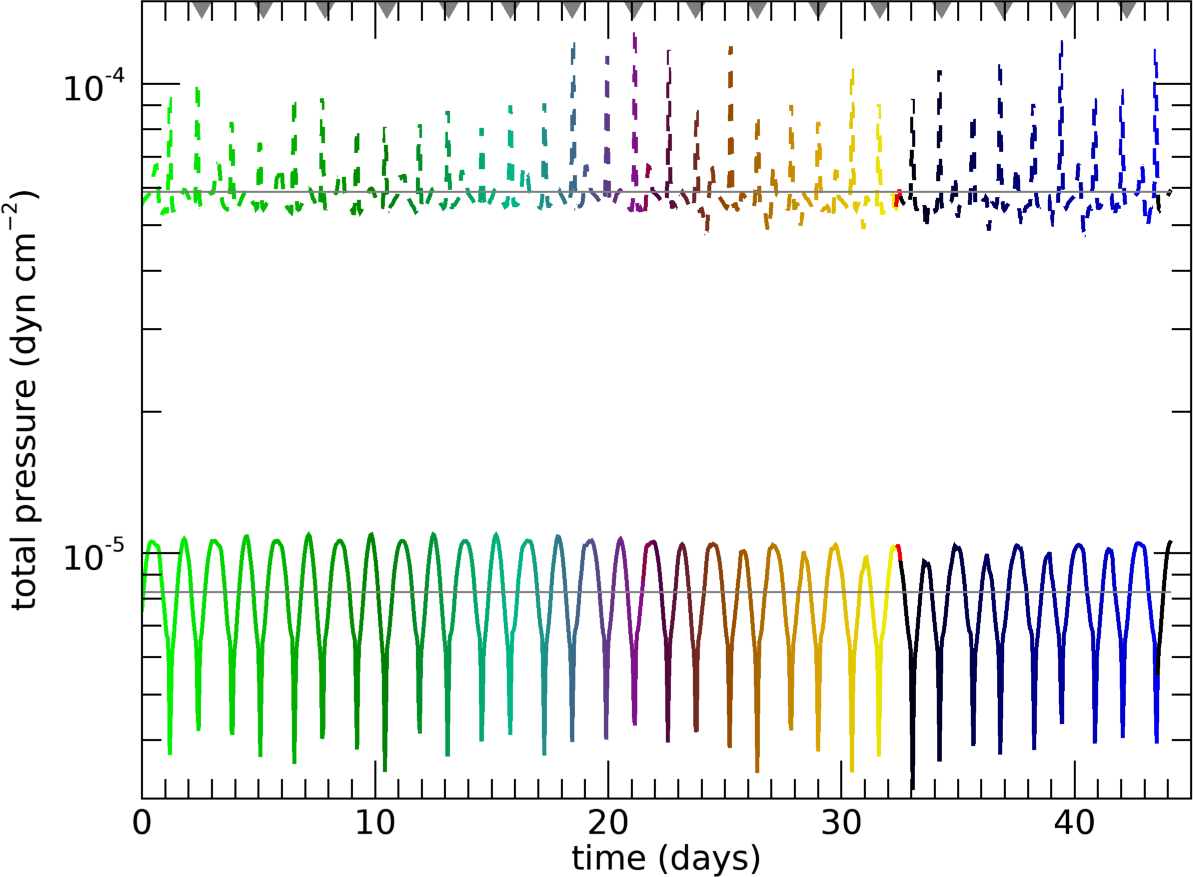}  
    \includegraphics[width=.48\textwidth]{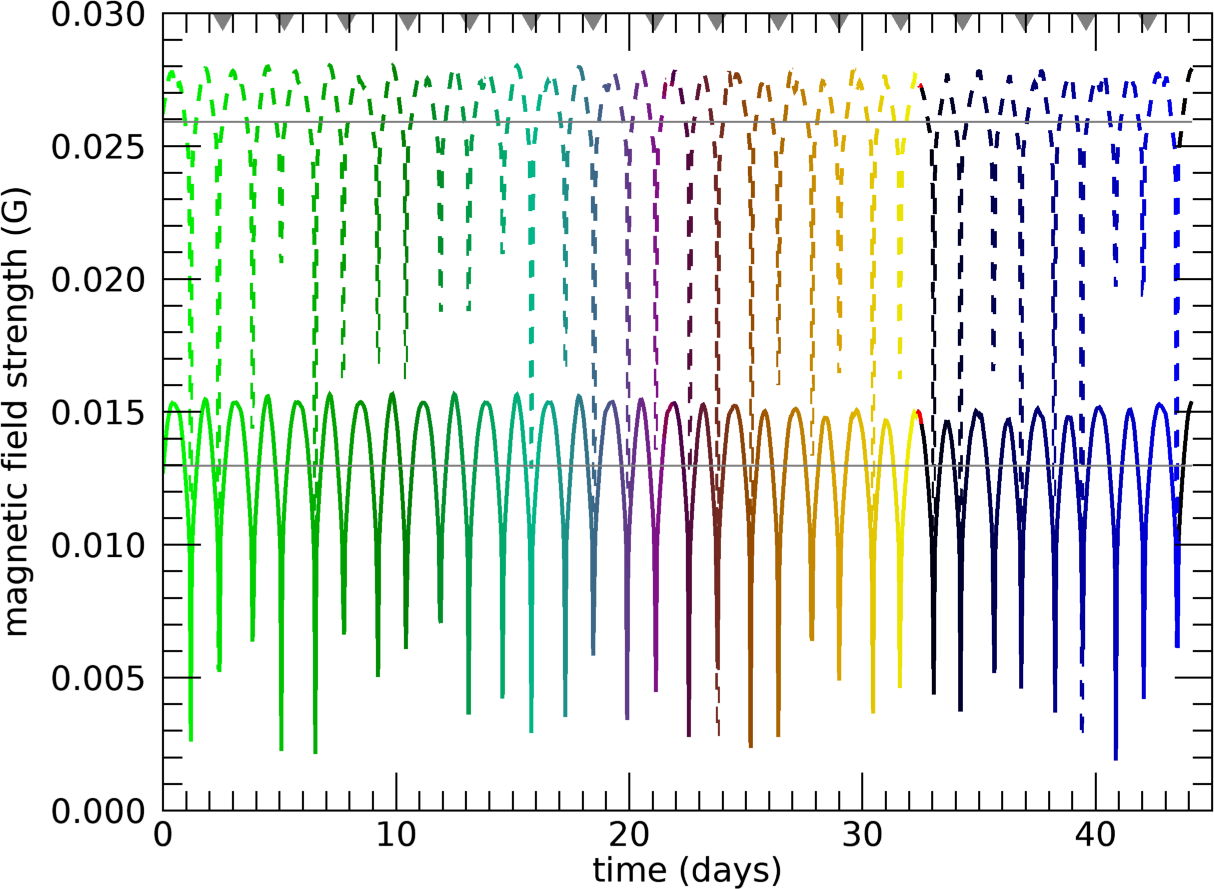}  
    \caption{The stellar wind proton density,  velocity, total pressure (sum of ram, including planet's own orbital motion,  thermal and magnetic pressures) and total magnetic field strength at the orbit of GJ\,436\,b.  The {solid} lines are for Model I and the {dashed} lines for Model II. Colour represents the time evolution and follows the colour scheme shown in Figure \ref{fig.pl_trajectory}. The horizontal grey lines are the time average of the stellar wind properties over one stellar rotation period (about $16$ orbital periods of the planet). {The grey arrows in the upper horizontal axes indicate approximate times of mid-transits. In the upper panels, the crosses indicate the wind values derived in the work of \citet{2016A&A...591A.121B} and they are plotted at the mid-transit times. Note how they overlap really well with the values derived in our Model I, indicating that Model I provides a better description of the wind of GJ\,436.}}
    \label{fig.wind}%
\end{figure*}

As presented in Section \ref{sec.ref}, in the case of GJ\,436\,b, which has a highly misaligned orbit, the time it takes for the planet to be at the same substellar point is a  complex combination of orbital and rotation periods, as well as the obliquity of the orbit. Because of this, the local condition of the stellar wind is constantly changing. Figure \ref{fig.wind} shows the stellar wind proton density, velocity, total pressure (ram + thermal + magnetic) and magnetic field strength, all computed at the orbit of GJ\,436\,b. The {solid} lines are for Model I, and the {dashed} lines for Model II. Colour represents the time evolution following the same colour scheme as shown in Figure \ref{fig.pl_trajectory}. Note that to compute the ram pressure, we use the relative velocity between the stellar wind velocity $\vec{u}$ and the Keplerian motion of the planet $\vec{v}_K$: $\Delta \vec{u} = \vec{u} - \vec{v}_{K}$. For GJ\,436\,b, the Keplerian velocity is $\vec{v}_K = 118 ( \sin \Psi \hat{\theta} +  \cos \Psi \hat{\varphi})~$km~s$^{-1}$, where we adopted a stellar mass of $\simeq 0.45~M_\odot$ and an orbital distance of $14.56~R_\star \simeq 0.028$~au \citep{2022MNRAS.514...77M}.

The range of wind densities at the orbit of the planet is $[0.12,5.0]\times 10^3 ~{\rm cm}^{-3}$ for Model I and $[2.8, 81]\times 10^3 ~{\rm cm}^{-3}$ for Model II. Their time averages over one stellar rotation period  are $510$ and $6200 ~{\rm cm}^{-3}$ for Models I and II, respectively (horizontal grey lines shown in Figure \ref{fig.wind}).  Similarly, for wind velocities, we found that the range of values that the planet experience are $[30, 770]~$km~s$^{-1}$ and $[220, 830]$~km~s$^{-1}$ for Models I and II, respectively, with averages of $510$ and $640$~km~s$^{-1}$. When modelling the transmission spectroscopic transit of GJ\,436\,b, \citet{2016A&A...591A.121B} derived densities of  $2^{+2.2}_{-1.2}\times 10^{3}~{\rm cm}^{-3}$ and wind speeds of $85^{+6}_{-16}$~km~s$^{-1}$.
 {These observations are taken during transit, so the values they derived should be compared to our model values around mid-transit times. For this comparison, we added to the top panels of Figure \ref{fig.wind} crosses at the approximate mid-transit times at the same level of  ion density ($2\times 10^{3}~{\rm cm}^{-3}$)  and speed ($85$~km~s$^{-1}$) derived in the work of  \citet{2016A&A...591A.121B}. We see that these crosses  overlap really well with values of Model I (solid lines), but not with Model II (dashed lines), implying}  that Model I, with the lowest surface flux of Alfv\'en waves, better matches the spectroscopic transit values from \citet{2016A&A...591A.121B}.  Indeed, from these observations, the estimated mass-loss rate of $1.2^{+1.3}_{-0.75}\times 10^{-15}~M_\odot~\rm{yr}^{-1}$ \citep{2017MNRAS.470.4026V} better agree with the values derived in Model I ($1.1 \times 10^{-15}~M_\odot~\rm{yr}^{-1}$), but not with Model II ($2.5 \times 10^{-14}~M_\odot~\rm{yr}^{-1}$). Our chosen values of $S_{A,0} / B_\star$ correspond to non-thermal velocities $\xi \simeq 14$ and $45$~km~s$^{-1}$ at the transition region (see Section \ref{sec.model}). Recently, \citet{2023ApJ...950..124B} reported $\xi$ ranging from $\sim 5$ to $\sim 30$~km~s$^{-1}$ for stars with similar rotation periods and effective temperatures as GJ\,436. In this case too, the non-thermal velocities from Model I have a better agreement with the range reported in \citet{2023ApJ...950..124B}. Altogether, our derived mass-loss rates and non-thermal velocities suggest that the `canonical' value of $S_{A,0} / B_\star = 1.1 \times 10^{7}~\textrm{erg (cm}^{2}\textrm{\,s\,G)}^{-1}$, derived from solar wind models, might not be applicable to the winds of (old) M dwarf stars.

%ion density
%int, model 1: ave, min, max:       515.871      124.398      5076.69
%tecplot    1: ave, min, max:        551.34256       124.37983       5533.9699
%int, model 2: ave, min, max:       6233.98      2772.02      81528.9
%tecplot    2: ave, min, max:        6572.8005       2772.0011       92838.510

%Utot
%int, model 1: ave, min, max:       516.008      29.3543      773.244
%tecplot    1: ave, min, max:       503.693      8.12634      782.222
%int, model 2: ave, min, max:       642.262      217.829      831.518
%tecplot    2: ave, min, max:       634.260      204.705      831.604

The bottom right panel of Figure \ref{fig.wind} shows the stellar wind magnetic field strength at the orbit of GJ\,436\,b. We find an average stellar magnetic field strength of $0.013$~G  and $0.026$~G for Models I and II, respectively. We use these values to then compute the magnetic pressure ($B^2/8\pi$). We  also compute the ram pressure of the stellar wind ($\rho \Delta u^2$). Adding these two pressures to the thermal pressure, we obtain the total pressure of the stellar wind at the position of the planet, which is shown at the bottom left panel of Figure \ref{fig.wind}. We find average total pressures of $8.2 \times 10^{-6}$~dyn/cm$^{-2}$ and  $59 \times 10^{-6}$~dyn/cm$^{-2}$ for Models I and II, respectively. These values will be used in Section \ref{sec.spi}, when we compute the strength of star-planet interactions and the size of the magnetosphere of the planet. 

%ptot
%int, model 1: ave, min, max:   8.26563e-06  3.12572e-06  1.09535e-05
%tecplot    1: ave, min, max:   8.13436e-06  2.91537e-06  1.11035e-05
%int, model 2: ave, min, max:   5.88245e-05  4.80686e-05  0.000132647
%tecplot    2: ave, min, max:   5.88313e-05  4.77303e-05  0.000152073

% btot
%int, model 1: ave, min, max:     d   0.00187580    0.0156699
%tecplot    1: ave, min, max:     0.0128187   0.00138458    0.0157221
%int, model 2: ave, min, max:     0.0259154   0.00265763    0.0280401
%tecplot    2: ave, min, max:     0.0258026  0.000187708    0.0282044

%%%%%%%%%%%%%%%%%%%%%%%%%%%%%%%%%%%%%%%%%%%%%%%%%
\section{Stellar-wind--planet interactions}\label{sec.spi}
Because of its close proximity to the star, and its sub-Alfv\'enic motion, GJ\,436\,b can magnetically interact with its host star. Recently, \citet{2023AJ....165..146L} studied the presence of star-planet interaction signatures in GJ\,436 in UV spectroscopic observations. While they did not find line fluxes modulated with the orbital period of  GJ\,436\,b (traditionally used as evidence of star-planet interactions, e.g., \citealt{2023NatAs.tmp...65P}), they found an enhancement of low energy flares and a lack of more energetic flares in  GJ\,436, compared to other M dwarfs. They suggested that this could be caused by the presence of GJ\,436\,b, whereby the reduction of the number of high energy flares in GJ\,436 would be caused by the planet  triggering early flaring events, i.e.,  before enough energy would have been built up in a larger stellar flare. As a consequence,  high-energy flares in GJ\,436 would not have time to store energy and would be prematurely released, explaining the excess of low energy flares and lack of high energy flares in the system. In their analysis, they also derived a maximum power for the star-planet interaction, detected in the far ultraviolet lines, of $\sim 3\times 10^{24}$~erg~s$^{-1}$. Converting this number to a bolometric power, they found an upper limit of $\lesssim 10^{26}$~erg~s$^{-1}$ for the maximum power released in star-planet interactions in the GJ\,436 system. 

To compare with the derived power from \citet{2023AJ....165..146L}, in this section, we compute  the strength of star-planet interactions. Here, we use two different scenarios for this calculation. One scenario is based  on the idea that magnetic reconnection occurring between stellar and planetary magnetic field lines releases magnetic energy \citep{2008A&A...490..843J, 2009A&A...505..339L, 2010ApJ...720.1262V}. This energy can accelerate electrons, which travel towards the star. This idea has been used to explain anomalous hot spots at  stellar chromospheres \citep{2009A&A...505..339L}, as reported in observational works \citep[e.g.,][]{2005ApJ...622.1075S, 2009EM&P..105..373P, 2019NatAs...3.1128C}. The other scenario we also explore is based on the idea that a planet moving at sub-Alfv\'enic speed through the magnetised wind of its host star can trigger MHD Alfv\'en waves, which can travel towards the star along magnetic field lines \citep{2006A&A...460..317P,2013A&A...552A.119S, 2016ApJ...833..140S}. As they travel, these waves dissipate their energies, which are radiated away. 

We now proceed to estimate the power released in the star-planet interaction using the two scenarios mentioned above. To compute the strength of star-planet interactions in both scenarios, we start by calculating the (unsigned) Poynting flux of the stellar wind that impacts on the planet
\begin{equation}\label{eq.poy}
P =  \frac{|(\Delta \vec{u} \times \vec{B}) \times \vec{B}|}{4\pi} = \frac{B^2 \Delta u \sin\theta}{4\pi} \, .
\end{equation} 
In the previous expression, $B\sin \theta$ is the  component of the stellar wind magnetic field  that is perpendicular to $\Delta \vec{u}$, $\theta$ is the angle between $\Delta \vec{u}$ and $\vec{B}$ -- all these quantities are computed at the position of the planet, which lies at an orbital distance $a_{\rm orb} = 14.56~R_\star$. 
We find that the average Poynting fluxes computed at the orbit of GJ\,436\,b are $P \simeq 160$ and $650$~erg~cm$^{-2}$~s$^{-1}$ (or $0.1$ to $0.6$~W~m$^{-2}$) for Models I and II, respectively. In the context of planets moving at super-Alfv\'enic speeds, the Poynting flux can be used to estimate the amount of  power dissipated in the magnetosphere of a planet, that can then power electron-cyclotron radio emission \citep[e.g.][]{2019MNRAS.488..633V}. In the case of GJ\,436\,b, though, our wind models  indicate that the motion is sub-Alfv\'enic (c.f.~Section \ref{sec.model}) and we use the Poynting flux to compute the energy released in the Alfv\'en wing scenario and the reconnection scenario. Note that the Poynting flux only depends on the characteristics of the wind at a certain orbital distance and does not depend on the properties of the planet, such as its size or magnetic field strength, contrary to the power released in star-planet interactions, as we see below.

% poynting flux
%int, model 1: ave, min, max:       163.108     0.906704      606.955
%tecplot    1: ave, min, max:       166.934     0.205199      609.485
%int, model 2: ave, min, max:       653.841      3.27953      1263.54
%tecplot    2: ave, min, max:       648.829    0.0680350      1263.98

%%%%%%%%%%%%%%%%%%%%%%%%%%%%%%%%%%%%%
\subsection{Scenario 1: Alfv\'en wings}
\begin{figure}[!h]
    \centering
    \includegraphics[width=.48\textwidth]{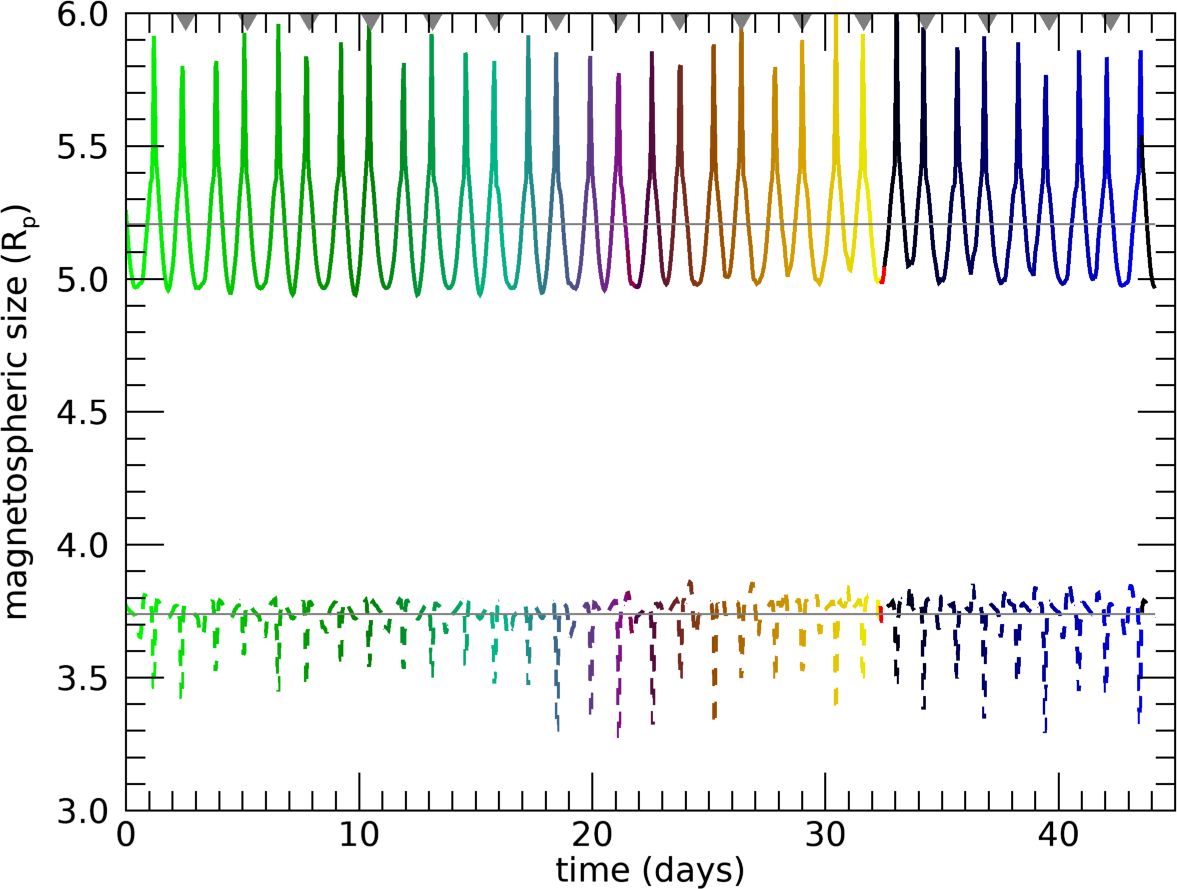} \\
    \includegraphics[width=.48\textwidth]{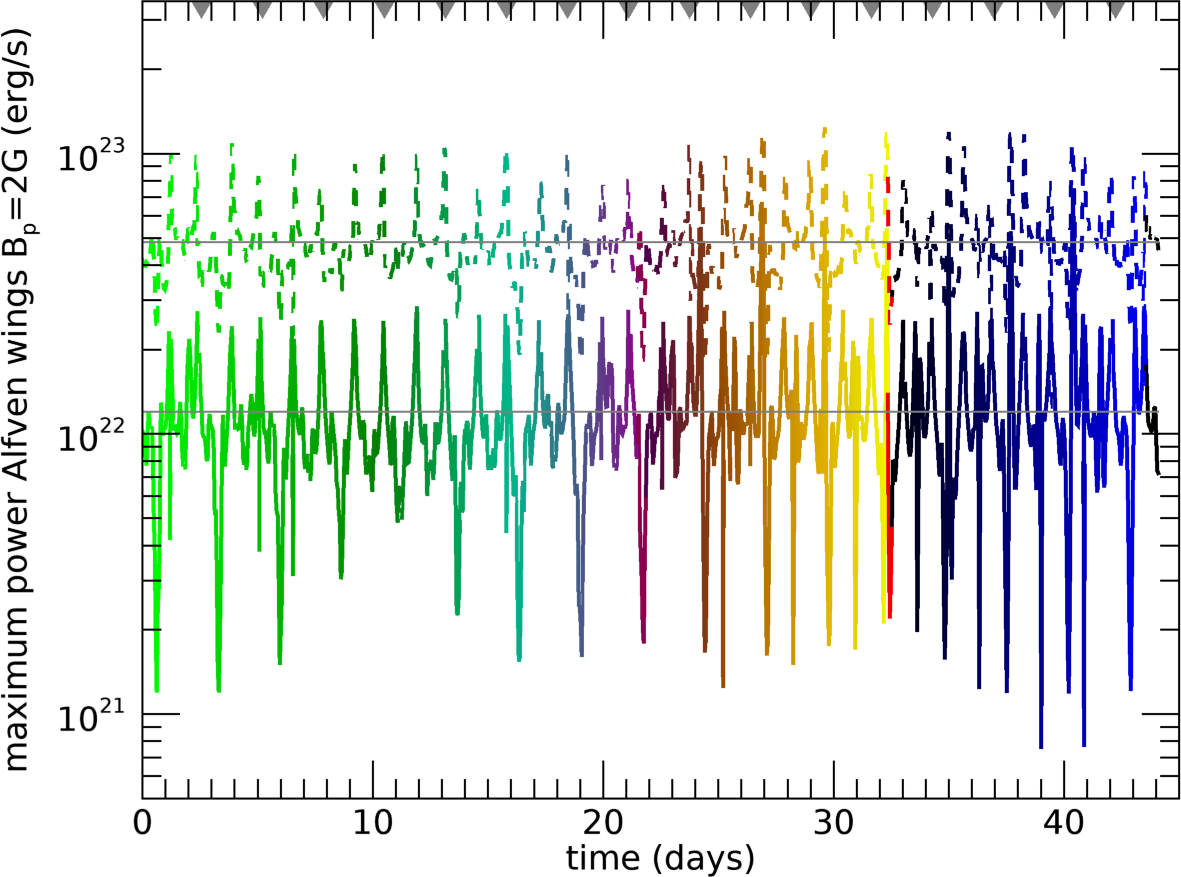}\\
    \includegraphics[width=.48\textwidth]{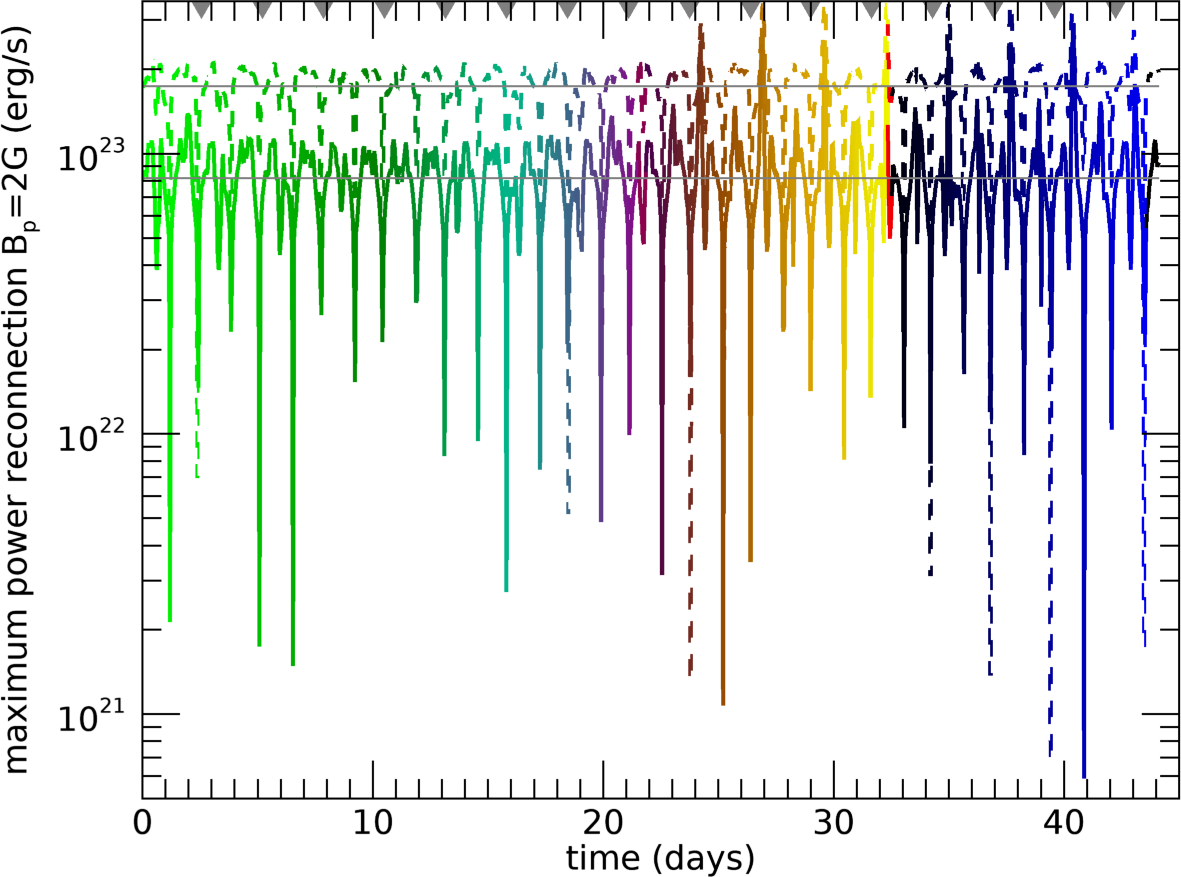} 
    \caption{Strength of the stellar-wind planet interactions computed in this work. Top: the magnetospheric size of the planet, assuming a planetary magnetic field strength of 2~G at the pole (i.e., 1~G at the equator). Middle: the maximum dissipated power of Alfv\'en waves travelling towards the star \citep{2013A&A...552A.119S}. Bottom:  the maximum dissipated power from the star-planet magnetic reconnection scenario \citep{2009A&A...505..339L}. The {solid} lines are for Model I, and the {dashed} lines for Model II. The colour scheme follows the one adopted in Figure \ref{fig.pl_trajectory}.  {The grey arrows in the upper horizontal axes indicate approximate times of mid-transits.}}
    \label{fig.fluxes}%
\end{figure}

A planet moving at sub-Alfv\'enic speed through the magnetised wind of its host star can trigger MHD Alfv\'en waves, which can travel towards the star along magnetic field lines. As they travel, these waves can then dissipate their energies and the maximum power that is radiated away can be estimated by integrating the Poynting flux of the stellar wind over the cross-section of two Alfv\'en `wings'. In the limit of sub-Alfv\'enic speeds, with Alfv\'en-Mach number $M_A = \Delta u/v_A < 1$, the dissipated power is \citep{2013A&A...552A.119S}
 \begin{equation}
 \mathcal{P}_{\rm wg} \simeq P (2 \pi r_M^2) \alpha^2 M_A \sin \theta  =   \frac{ \Delta {u}^2 {B} \sin^2 \theta }{2} r_M^2 \alpha^2   \sqrt{4\pi\rho}
 \, .
 \end{equation}
Here, $\alpha$ is a dimensionless parameter related to the conductive properties of the planet and it can be understood as the strength of the interaction; here, we assume $\alpha =1$, thus deriving the maximum power $ \mathcal{P}_{\rm wg}$. In the expression above, the planet is assumed to be magnetised, with a magnetospheric size $r_M$. Through pressure balance between the magnetic pressure of the planet and the total pressure of the stellar wind (i.e., sum of the magnetic, thermal and ram pressures, see bottom left panel of Figure \ref{fig.wind}), we have \citep[e.g.,][]{2014MNRAS.438.1162V} 
\begin{equation}\label{eq.r_M}
 \frac{r_M}{R_p} \simeq f \left[  \frac{(B_p/2)^2/(8 \pi)}{ p_{\rm tot} (a_{\rm orb})} \right]^{1/6}.
\end{equation}
where $B_{p}$ is the polar field strength of the planet's dipolar magnetic field (i.e., twice the value of the equatorial field strength) and $f\simeq 2^{2/6}$ is a correction factor used to account for the effects of currents \citep[e.g.][]{2004pssp.book.....C}. The value of $B_{p}$ for GJ\,436\,b is currently poorly constrained: \citet{2016A&A...591A.121B} derived an upper limit for the equatorial field strength of $\sim 3$~G, consistent with the upper limit estimates of $\sim 10$~G from \citet{2023AJ....165..146L}. Assuming $B_{p}=2$~G leads to an average magnetosphere size of $5.2~R_p$ and $3.7~R_p$, for stellar wind Models I and II, respectively, as seen on the top panel of Figure \ref{fig.fluxes}. The maximum power that is radiated by Alfv\'en waves travelling towards the star is shown on the middle panel of this figure, where we see on average that the radiated power is $\mathcal{P}_{\rm wg} \simeq 1.2 \times 10^{22}$~erg~s$^{-1}$ and $4.8\times 10^{22}$~erg~s$^{-1}$ for Models I and II, respectively.  It is worth noting that the radiated power depends on the assumed magnetic field strength of the planet indirectly through $r_M$. As $r_M \propto B_p^{1/3}$, we have that $\mathcal{P}_{\rm wg} \propto B_p^{2/3}$. Thus for a 20-G polar field strength (i.e., 10 G at the equator), the  radiated power increases by $10^{2/3}$, i.e., nearly a factor of 5, resulting in $ \mathcal{P}_{\rm wg} \simeq 5.6\times 10^{22}$~erg~s$^{-1}$ and $22\times 10^{22}$~erg~s$^{-1}$ for Models I and II, respectively. 

%R_M
 %int, model 1: ave, min, max:       5.20680      4.93982      6.08814
%tecplot    1: ave, min, max:        5.2221891       4.9286346       6.1591539
%int, model 2: ave, min, max:       3.73917      3.26002      3.86065
%tecplot    2: ave, min, max:        3.7394543       3.1863582       3.8651912

 %Flux Saur AW
%int, model 1: ave, min, max:   1.20276e+22  7.51691e+20  7.10987e+22
%tecplot    1: ave, min, max:    1.2716861e+22   4.5219853e+19   7.1857497e+22
%int, model 2: ave, min, max:   4.83866e+22  1.20735e+22  1.24016e+23
%tecplot    2: ave, min, max:    4.8800454e+22   1.4399140e+20   1.2410704e+23

%10G Saur
%int, model 1: ave, min, max:   5.58273e+22  3.48905e+21  3.30012e+23
%tecplot    1: ave, min, max:    5.9026441e+22   2.0989200e+20   3.3353300e+23
%int, model 2: ave, min, max:   2.24591e+23  5.60406e+22  5.75632e+23
%tecplot    2: ave, min, max:    2.2651166e+23   6.6834888e+20   5.7605389e+23

%%%%%%%%%%%%%%%%%%%%%%%%%%%%%%%%%%%%%%%%%%
\subsection{Scenario 2: Star-planet magnetic reconnection}      
Magnetic reconnection occurs when magnetic field lines of different polarities interact with each other. Magnetic reconnection has been considered in the case of star-planet interactions by several authors \citep[e.g.][]{2008A&A...490..843J,2009A&A...505..339L, 2010ApJ...720.1262V}. While \citet{2008A&A...490..843J, 2010ApJ...720.1262V} considered reconnection events to power radio emission from exoplanets, \citet{2009A&A...505..339L} studied how reconnection events could trigger chromospheric hot spots. We can estimate the power released in the reconnection by integrating the Poynting flux $P$ (Equation \ref{eq.poy}) over the cross-section of the planet's magnetosphere \citep{2009A&A...505..339L}
 \begin{equation}\label{eq.nuccio}
 \mathcal{P}_{\rm rec} = \gamma_{\rm rec}  P (\pi R_M^2)  = \frac14 \gamma_{\rm rec}  r_M^2 {B^2 \Delta u \sin\theta} \, , 
 \end{equation}
 where $ \gamma_{\rm rec}$ is an efficiency of the reconnection related to the angle between the stellar and planetary field lines. Even though in the GJ\,436 system the planet will move through different orientations of the stellar wind magnetic field (e.g., going through closed and open field lines), and thus $\gamma_{\rm rec}$ would differ at different points in the orbit, for simplicity here we take its maximum value of $ \gamma_{\rm rec}=1$, so that our power estimates $ \mathcal{P}_{\rm rec}$ are upper limits. The results of this scenario are shown in the bottom panel of Figure \ref{fig.fluxes}, where we see that the average powers released in the reconnection scenario are $ \mathcal{P}_{\rm rec} \simeq 8\times 10^{22}$~erg~s$^{-1}$ and $17\times 10^{22}$~erg~s$^{-1}$ for Models I and II, respectively, where we assumed a planetary polar field strength of 2~G. Similarly to the previous scenario we considered, the power released in the reconnection scenario also depends on the planetary magnetic field as $\mathcal{P}_{\rm rec} \propto B_p^{2/3}$, thus for a 20-G field, we find an increase in dissipated power of nearly a factor of 5, reaching average values of $ \mathcal{P}_{\rm rec} \simeq 38\times 10^{22}$~erg~s$^{-1}$ and $80\times 10^{22}$~erg~s$^{-1}$ for Models I and II, respectively.

%flux reconnection      
%int, model 1: ave, min, max:   8.17043e+22  5.89983e+20  2.89095e+23
%tecplot    1: ave, min, max:    8.3716753e+22   1.3988572e+20   2.9052466e+23
%int, model 2: ave, min, max:   1.74268e+23  6.73300e+20  3.47779e+23
%tecplot    2: ave, min, max:    1.7283159e+23   1.4316130e+19   3.4790195e+23

%10G eq field
%int, model 1: ave, min, max:   3.79238e+23  2.73846e+21  1.34186e+24
%tecplot    1: ave, min, max:    3.8857877e+23   6.4929211e+20   1.3484962e+24
%int, model 2: ave, min, max:   8.08882e+23  3.12518e+21  1.61425e+24
%tecplot    2: ave, min, max:    8.0221323e+23   6.6449593e+19   1.6148179e+24

\subsection{Interconnecting magnetic loop between star and planet}
Here, we calculated the power originating from star-planet interactions mediated by the stellar wind, using the two scenarios discussed before. There is still a third model that we explore in the Appendix, namely that of an interconnecting loop that extends from the stellar surface to the planetary orbit \citep{2013A&A...557A..31L}. For similar planetary field strengths, the interconnecting loop scenario produces maximum powers that are about three to four orders of magnitude larger than the other scenarios. This scenario assumes that the stellar magnetic field remains closed out to the orbital distance of the planet, which is not seen in our wind models. 

For the three scenarios explored here (in Section \ref{sec.spi} and in the Appendix), the predicted powers show peaks and troughs, which are due to the planet crossing the magnetic equator of the star (when the field changes polarity). To compute the visibility of this interaction, one would need to take into account inclination of the rotation axis of the star ($35.7^{+5.9}_{-7.6}$~deg, \citealt{2022A&A...663A.160B}), the timing when the planet is hidden behind the stellar disk, and possible phase lags.

\section{{Discussion and} Conclusions}\label{sec.conclusion}

In this paper, we performed 3D numerical simulations of the wind of GJ\,436. Our model incorporates the recently reconstructed radial magnetic field of GJ\,436 \citep{2023arXiv230615391B} that shows that, in March-June 2016, GJ\,436 presented a large-scale field that largely resembles an aligned dipole, with a maximum (absolute) field strength of $\sim 27~$G (Figure \ref{fig.zdi}). 
Stellar outflows, in the form of winds and coronal mass ejections, shape the space weather environment around planets. From our simulations, we derived the space weather conditions around GJ\,436\,b, a warm-Neptune orbiting at a distance of $0.028~\rm{au}$ from its host star. The main unknown in our wind models is the energy flux of the Alfv\'en waves that heat and accelerate the stellar wind. We ran two models with different values of the Alfv\'en wave fluxes to magnetic field ratio: the `canonical' value of $S_{A,0} / B_\star = 1.1 \times 10^{7}~\textrm{erg (cm}^{2}\textrm{\,s\,G)}^{-1}$, commonly used in solar wind simulations, and another value that is one order of magnitude smaller. These models were named Model I (with the lower flux) and Model II (with the higher flux), as can be seen in Figure \ref{fig:3d}. With the exception of the Alfv\'en wave properties, the remaining inputs of our models are well constrained by observations, including the large-scale surface field of the star.  With our models, we derive stellar wind mass loss rates of $1.1 \times 10^{-15}~M_\odot~\rm{yr}^{-1}$ and $2.5 \times 10^{-14}~M_\odot~\rm{yr}^{-1}$, for Models I and II, respectively. {As discussed further below, we find that Model I is a better representation of the wind of GJ\,436.}

In addition to the global properties of the stellar wind, our models also allow us to derive the conditions of the stellar wind at the orbit of GJ\,436\,b. Even though at the observed epoch the stellar magnetic field, and thus the stellar wind, was approximately axi-symmetric about the rotation axis of the star, GJ\,436\,b has a nearly polar orbit. This means that during one  orbit, the planet probes regions of open field lines and closed field lines. Taking into account the spin-orbit angle, we calculated the position of the planet during consecutive orbits: during one stellar rotation ($\sim 44$ days) the planet orbits about $\sim 16$ times, experiencing different space weather conditions at each orbit (see Figures \ref{fig.pl_trajectory} and \ref{fig.wind}). There are two main sources of variation in these conditions: one relates to the choice of wind model (lower versus higher Alfv\'en wave fluxes) and the other relates to the high obliquity of the orbit of the GJ\,436\,b. For example, the many peaks and troughs we see in the local wind properties (Figure \ref{fig.wind}) are associated to the planet crossing the  stellar equatorial plane. 

Regardless of the two wind models we choose, our 3D simulations indicate that the motion of GJ\,436\,b through the stellar wind is sub-Alfv\'enic, which implies that the planet has a direct magnetic {connectivity} with its host star. In a second part of our study, we investigated the power released in star-planet interactions, mediated by the magnetised stellar wind. We first computed the Poynting fluxes of the stellar wind that impacts on GJ\,436\,b to be $P \simeq 160$ and $650$~erg~cm$^{-2}$~s$^{-1}$ (or $0.1$ to $0.6$~W~m$^{-2}$) for Models I and II, respectively. Next, we computed the maximum energy that is available to power sub-Alfv\'enic star-planet interactions through two different scenarios: the Alfv\'en wing scenario (in which perturbations caused in the stellar magnetic field line by the planet travel towards the star as MHD waves, irradiating their energies, \citealt{2013A&A...552A.119S}) and the magnetic reconnection scenario (in which magnetic reconnection between stellar and planetary magnetic fields release energy, \citealt{2009A&A...505..339L}). For both scenarios, we assume that the planet is magnetised with an equatorial dipolar magnetic field of 1G, leading to a magnetospheric size of about  $5.2~R_p$ and $3.7~R_p$, for stellar wind Models I and II, respectively. 
Our estimated maximum powers released through star-planet interactions are of the order of $10^{22}$ to $10^{23}$~erg~s$^{-1}$ (approximately $10^{-10}$ to $10^{-9}$ of the stellar bolometric luminosity), with the largest values associated to the reconnection scenario. It is interesting to note that the powers we derived are not too sensitive to the choice of Alfv\'en wave surface flux in the wind model  (the largest unknown in our  wind models) -- a change of one order of magnitude in the surface flux of Alfv\'en waves leads to average powers that are a factor of 4 different for the Alfv\'en wing scenario and a factor of 2 in the reconnection scenario only (Figure \ref{fig.fluxes}).

We note here that the planetary magnetic field is an assumption in our calculations and that the power released in the two star-planet interaction scenarios we considered increase with $B_p^{2/3}$.   Thus, increasing the magnetic field of the planet by a factor of $10$ (i.e., to an equatorial field of 10~G) leads to an increase in the estimated powers by $10^{2/3}$, i.e., nearly a factor of 5. We could keep increasing the choice of planetary field strength until our estimated powers increase from $\sim 10^{23}$~erg~s$^{-1}$ to $10^{26}$~erg~s$^{-1}$, the latter of which is the  maximum power derived by \citet{2023AJ....165..146L} due to star-planet interactions in the GJ\,436 system. To match the observed upper limit, the maximum planetary magnetic field would need to be unreasonably high ($\lesssim10^{3 \times {3/2}}\simeq 3\times 10^4$ ~G). Because of this, unfortunately, we cannot  derive any meaningful upper limit to the planetary magnetic field. 

{According to the models studied here, GJ\,436\,b orbits mostly within the Alfven surface, i.e., it has a sub-Alfvenic orbital motion, except for small portions of the orbit around the magnetic equator, where the orbit is outside the Alfven surface. When a planet moves from sub- to super-Alfvenic motion, it is likely that different types of star-planet interactions can take place. An immediate effect should be seen by the formation of a bow shock around the planet, when it moves from sub- to super-Alfvenic motion \citep[e.g.][]{2011MNRAS.414.1573V}.  \citet{2014ApJ...790...57C} studied the sub-/super-Alfvenic transition in terms of Joule heating in the upper atmosphere of planets, while \citet{2018ApJ...858..105K} and \citet{2020A&A...636L..10K} demonstrated that planets orbiting in the varying magnetic field of their host stars can experience induction heating also in their interiors. Additionally, it is also expected that magnetic connectivity can switch between on and off states -- this can be due to variations along one planetary year (e.g., the planet moving in and out of the Alfven surface), or variations caused to the Alfven surface because of the evolution of the surface magnetic field \citep{2010MNRAS.406..409F}. The latter fact has been attributed as the cause of the disappearance of planet-induced chromospheric modulation in planet-hosting stars \citep{2008ApJ...676..628S}, emphasising the importance of contemporaneous observations of stellar magnetism and search for signatures of star-planet interactions.  }

We also explored a third star-planet interaction scenario in the Appendix of this paper -- in this scenario, a magnetic loop connecting the star and the planet would suffer a rupture after being stretched by the planet \citep{2013A&A...557A..31L}. This scenario has been recognised as the one that releases the largest star-planet interaction powers, compared to the other scenarions \citep{2019NatAs...3.1128C, 2022MNRAS.512.4556S}. However,  this scenario is not applicable to the case where the star-planet interaction is mediated by a stellar wind that distorts the stellar magnetic field from a potential state. In this case, in the calculations shown in the Appendix, a potential field model for the stellar magnetic field is assumed (without a stellar wind). Using this scenario, we demonstrate in the Appendix that a planetary field strength of $\sim 6$~G would reproduce the powers found by \citet{2023AJ....165..146L}, agreeing with these authors'  estimates.

In our work, we adopted two different values of the Alfv\'en wave flux that drives the stellar wind. One way to constrain the best choice of surface Alfv\'en wave flux, and thus decide which of the wind models is more realistic, is to compare the mass-loss rate and wind properties we predict with the values obtained from models of the transmission spectroscopic transit of GJ\,436\,b from \citet{2016A&A...591A.121B}. Comparing the wind velocities and densities that we derived to those derived in \citet[][{shown as grey crosses in the upper panels of Figure \ref{fig.wind}}]{2016A&A...591A.121B}, we found that our Model I, with the lowest surface flux of Alfv\'en waves, better matches the spectroscopic transit values. Our derived mass-loss rate of $1.1 \times 10^{-15}~M_\odot~\rm{yr}^{-1}$ is compatible with the one derived from spectroscopic transits of $1.2^{+1.3}_{-0.75}\times 10^{-15}~M_\odot~\rm{yr}^{-1}$ \citep{2017MNRAS.470.4026V}. {In addition to this, our preference towards Model I is also supported by further empirical and theoretical works, as we explain below.

Empirically, for the Sun, it has been suggested that the non-thermal velocity $\xi$ of lines formed at the transition region are associated with the rms amplitude of the Alfv\'en waves \citep[e.g.,][]{2017ApJ...845...98O} as $\xi = \langle v_{\perp,0}^2 \rangle^{1/2} /\sqrt{2}$. If this can be extended to solar-like stars, this implies that by measuring $\xi$ one can constrain the wave properties of other stars. Recently, \citet{2023ApJ...950..124B} measured the non-thermal velocity amplitudes of a sample of 55 low-mass stars. Overall, they found an increase in $\xi$ with stellar rotation (or activity). In their sample, they have about 5 stars with similar masses, radii and effective temperature to GJ\,436, namely: GJ\,832, LTT\,2050, GJ\,3470, GJ\,176, GJ\,849. Three of these stars (GJ\,832, GJ\,176, GJ\,849) have a rotation period of about $40$ days, very similar to GJ\,436. Using the Si IV  line (1393 \AA), \citet{2023ApJ...950..124B} measured  $\xi$ ranging between $14.7$ and $17.8$~km~s$^{-1}$ for these objects. These  values are very similar to the non-thermal velocity of $14$~km~s$^{-1}$ from Model I.

Theoretically, it is usually assumed that the wave flux is given by a combination of the available turbulent energy flux inside a magnetic flux tube times the magnetic filling factor \citep[e.g.][]{2017ApJ...840..114C, 2018PASJ...70...34S}. The magnetic filling factor is smaller for slowly rotating stars \citep{2019ApJ...876..118S}, which are also usually older. Compared to the Sun, the turbulent energy flux is also smaller in M dwarfs, given that the energy flux emerging from convective motions (and thus eventually driving the waves) is proportional to the bolometric energy flux of star $\sigma T_{\rm eff}^4$ \citep[e.g.,][]{2018PASJ...70...34S}. \citet{2021ApJ...919...29S} performed a parametric study of wave-driven wind models of M dwarfs, showing that the wave amplitude in the lower corona is smaller for M dwarfs than for the Sun. 

In conclusion, both models \citep{2021ApJ...919...29S}  and empirical measurements \citep{2023ApJ...950..124B} of the wave amplitude  point  to lower values of wave fluxes for slowly rotating M dwarfs. Based on these arguments and the agreement of the local wind values we derive here with results from Ly-$\alpha$ transit modelling \citep{2016A&A...591A.121B, 2017MNRAS.470.4026V}}, we favour a lower flux of Alfv\'en waves to be more appropriate to the case of GJ\,436. This might imply that the `canonical' value of $S_{A,0} / B_\star = 1.1 \times 10^{7}~\textrm{erg (cm}^{2}\textrm{\,s\,G)}^{-1}$, derived from solar wind models, might not be applicable to the winds of old M dwarf stars.

{Our models indicate that the planet GJ\,436\,b experiences a vast range of stellar wind speeds and densities along its orbit. Because the wind shapes the atmospheric material that is being evaporated from the planet, we  expect variations in the interaction region between the escaping atmosphere and the stellar wind, and hence, variations in the large structures formed due to this interaction, such as comet-like tails. We see the effects of different stellar wind properties on the morphology of comet-like tails, for example, in 3D models investigating these interactions \citep[e.g.][]{2014MNRAS.438.1654V, 2021MNRAS.501.4383V, 2019ApJ...873...89M, 2021MNRAS.500.3382C, 2022ApJ...927..238R}. Time-dependent effects, such as coronal mass ejections, would also alter the morphology of comet-like tails \citep{2017ApJ...846...31C, 2022MNRAS.509.5858H}, similar to what we see when the solar wind interacts with the plasma tail of comets \citep{2007ApJ...668L..79V}. Because of the large range of stellar wind properties experienced by GJ\,436\,b, similar temporal effects are expected during its orbit. It is interesting to note, however, the remarkable stability of the wind properties at every consecutive mid-transit time (crosses in the upper panel of Figure \ref{fig.wind}). If the \textit{local} wind properties are relatively constant for a few hours, this would lead to a relatively stable interaction zone, potentially explaining why the Ly-$\alpha$ transit of GJ\,436\,b remains stable over time \citep{2019A&A...629A..47D}. 

For this stability to occur in transits observed over many years apart, the stellar magnetic field cannot vary substantially. To investigate this, we need further spectropolarimetric campaigns to reconstruct the large-scale field of the star, and thus study its evolution.} We note that GJ\,436 has an activity cycle with a period of about 7 to 8 years  \citep{2018AJ....155...66L, 2019A&A...629A..47D, 2023MNRAS.518.3147K, 2023AJ....165..146L}, which implies that the stellar magnetic field could vary (in strength and topology) substantially at different phases of the cycle. This temporal variation would affect the wind properties and space weather conditions around the planet \citep[e.g.,][]{2012MNRAS.423.3285V, 2016MNRAS.459.1907N, 2018ApJ...864..125F, 2019MNRAS.485.4529K}, and, consequently, the released powers in star-planet interactions  \citep[e.g.,][]{2022MNRAS.512.5067K}. Ideally, observations that probe star-planet interactions mediated by the stellar  wind/magnetic field (e.g., planetary radio emission, chromospheric hot-spots, interaction of evaporated atmospheres with stellar winds) and the magnetic map used in stellar wind models should be conducted contemporaneously.

\begin{acknowledgements}
AAV thanks  P.~Loyd and A.~Lanza for useful discussions in the preparation of this paper. {The authors thank the anonymous reviewer for their constructive recommendations and suggestions.}  This project has received funding from the European Research Council (ERC) under the European Union's Horizon 2020 research and innovation programme (grant agreement No 817540, ASTROFLOW; grant agreement No 947634,  {\sc Spice Dune}; grant agreement No 740651, NewWorlds). RF acknowledges support from the United Arab Emirates University (UAEU) startup grant number G00003269. This work has been carried out within the framework of the NCCR PlanetS supported by the Swiss National Science Foundation under grants 51NF40$\_$182901 and 51NF40$\_$205606. We acknowledge funding from the French National Research Agency (ANR) under contract number ANR-18-CE31-0019 (SPlaSH). We thank SURF (www.surf.nl) for the support in using the National Supercomputer Snellius. This work used the BATS-R-US tools developed at the University of Michigan Center for Space Environment Modeling and made available through the NASA Community Coordinated Modeling Center. 
\end{acknowledgements}

\bibliographystyle{aa}
\bibliography{/Users/avidotto/Work/artigos/00bibtex-list,extra_ref}

\appendix

%%%%%%%%%%%%%%%%%%%%%%%%%%%%%%%%%%%%%%%%%%%%%%%%%%
\section{Calculation of orbital trajectories for a misaligned orbit}\label{sec.ref}
\begin{figure}[!b]
    \centering
    \includegraphics[width=.3\textwidth]{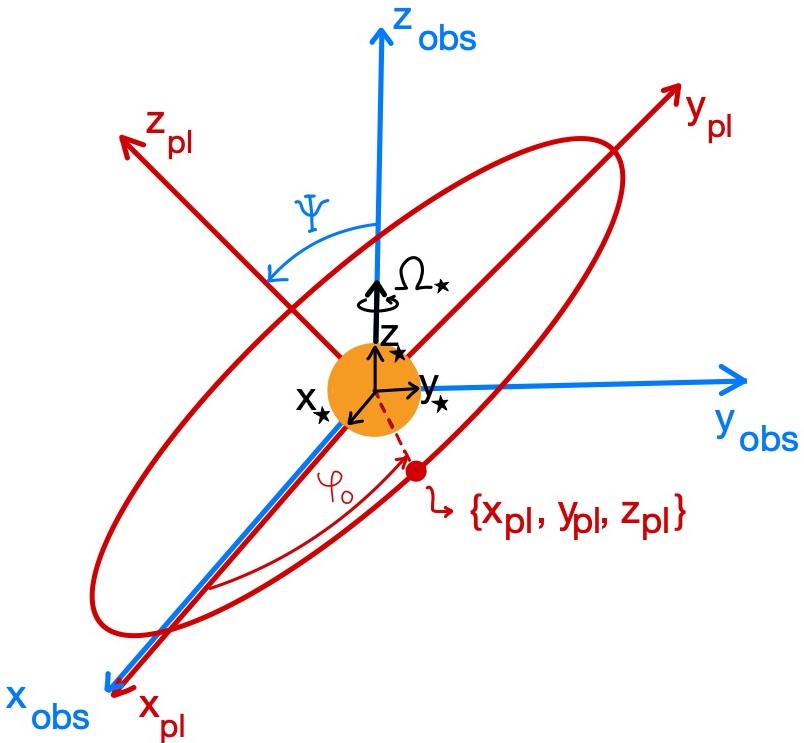} 
    \caption{Reference frames used in the coordinate transformation, {with the star given by the orange circle, the planet indicated by the red circle, and its orbital path shown in red. The planet's reference frame $\{x_{\rm pl}, y_{\rm pl}, z_{\rm pl}\}$ in red contains the  orbital plane of the planet $x_{\rm pl}y_{\rm pl}$ and orbital spin axis along $z_{\rm pl}$. The observer's inertial reference frame $\{x_{\rm obs}, y_{\rm obs}, z_{\rm obs}\}$ (in blue) is centred on the star, with the observer located  at  $+x_{\rm obs}$, and the stellar rotation axis aligned along $z_{\rm obs}$. The reference frame $\{x_\star, y_\star, z_\star\}$ (in black) co-rotates with the star. The spin-orbit angle is $\Psi$.}}
    \label{fig.ref}%
\end{figure}

To calculate the orbital trajectories in the reference frame of the star, we compute two coordinate transformations: one from the orbital reference frame to the observer's frame, and another from the observer's frame to the stellar frame. These frames are illustrated in Figure \ref{fig.ref}.

We start from the orbital reference frame $\{x_{\rm pl}, y_{\rm pl}, z_{\rm pl}\} $, such that the $x_{\rm pl}y_{\rm pl}$-plane contains the orbit of the planet and the orbital axis is along $z_{\rm pl}$. The origin of this reference frame is  centred on the star. In this reference frame, the (circular) planetary motion is simply described as
\begin{equation}\label{eq.orb1}
\begin{bmatrix}
 x_{\rm pl} \\ y_{\rm pl}\\ z _{\rm pl}
\end{bmatrix}
= 
\begin{bmatrix}
R_{\rm orb} \cos \varphi_{\rm pl} \\
R_{\rm orb} \sin \varphi_{\rm pl} \\
0
\end{bmatrix} \, , 
\end{equation}
where the phase of the orbit is $\varphi_{\rm pl} = \Omega_{\rm orb}t + \varphi_0$, with $t$ describing the time and $\Omega_{\rm orb}=2\pi/P_{\rm orb}$ the orbital rotation rate. Here, we assume a non-null initial phase $\varphi_0$, which we computed later in this section to find the location of GJ\,436\,b at zero phase of the surface magnetic map, {which corresponds to the first spectropolarimetric observation as reported in \citet{2023arXiv230615391B}}. 

Now we take the observer's (inertial) reference frame $\{x_{\rm obs}, y_{\rm obs}, z_{\rm obs}\} $, whose origin is also  centred on the star, and  $z_{\rm obs}$ is aligned with the stellar rotation. {The observer is located at positive $x_{\rm obs}$ and} we assume that $x_{\rm obs} \parallel x_{\rm pl}$. The angle between the orbital spin and the stellar spin axis is the spin-orbit angle $\Psi$, thus, the angle between $z_{\rm obs}$ and $z_{\rm pl}$ is also $\Psi$. In this reference frame, the orbital motion of the planet is described as
\begin{equation}\label{eq.obs}
\begin{bmatrix}
 x_{\rm obs} \\ y_{\rm obs}\\ z _{\rm obs}
\end{bmatrix}
= 
\begin{bmatrix}
1 & 0& 0 \\
0 & \cos \Psi & \sin \Psi \\
0 & -\sin \Psi & \cos \Psi
\end{bmatrix}
\begin{bmatrix}
x_{\rm pl} \\ y_{\rm pl}\\ z _{\rm pl}
\end{bmatrix} \, ,
\end{equation}
where we performed an anticlockwise rotation around the axis $x_{\rm obs} \equiv x_{\rm pl}$ by an angle $\Psi$. {According to our reference frame, the mid-transit occurs when the planet is at $\{ x_{\rm obs},  y_{\rm obs}, z_{\rm obs} \} =\{ R_{\rm orb}, 0,0\}$.}

The reference frame co-rotating with the star $\{x_\star, y_\star, z_\star\} $ is also centred on the star, with the stellar spin axis along $z_\star$, which is parallel to $z _{\rm obs}$. To transform from the observer's reference frame to the stellar co-rotating frame, we perform a clockwise rotation around the axis $ z_\star\equiv z_{\rm obs} $ by an angle $\Omega_\star t$
\begin{equation}\label{eq.star}
\begin{bmatrix}
 x_\star \\ y_\star \\ z_\star
\end{bmatrix}
= 
\begin{bmatrix}
\cos (\Omega_\star t) & -\sin (\Omega_\star t) &0\\
 \sin (\Omega_\star t) & \cos (\Omega_\star t) &0\\
0 & 0& 1 
\end{bmatrix}
\begin{bmatrix}
 x_{\rm obs} \\ y_{\rm obs}\\ z _{\rm obs}
\end{bmatrix} \, .
\end{equation}

Therefore, to derive the orbital path of the planet in the reference frame of the star, we substitute (\ref{eq.orb1}) and (\ref{eq.obs}) into (\ref{eq.star}), thus obtaining
\begin{equation}\label{eq.finalx}
\frac{ x_\star}{R_{\rm orb} }= \cos (\Omega_{\rm orb}t + \varphi_0) \cos (\Omega_\star t)    - \sin (\Omega_{\rm orb}t + \varphi_0) \sin (\Omega_\star t)  \cos \Psi 
\end{equation}
\begin{equation}\label{eq.finaly}
\frac{y_\star}{R_{\rm orb} } = \cos (\Omega_{\rm orb}t + \varphi_0) \sin (\Omega_\star t)  + \sin (\Omega_{\rm orb}t + \varphi_0) \cos (\Omega_\star t) \cos \Psi  
\end{equation}
\begin{equation}\label{eq.finalz}
\frac{ z_\star}{R_{\rm orb} }= -\sin (\Omega_{\rm orb}t + \varphi_0) \sin \Psi  \, .
\end{equation}
 Figure \ref{fig.pl_trajectory} shows the trajectories of the planet GJ\,436\,b as seen in the reference frame of the star during one rotational period of the star (about 44~days)

Note that, in the case of aligned systems, $\Psi=0$ and we have
\begin{eqnarray}
{ x_\star}= R_{\rm orb}\cos (\Omega_{\rm orb}t + \varphi_0+\Omega_\star t) \\
{y_\star} = R_{\rm orb}\sin (\Omega_{\rm orb}t + \varphi_0 + \Omega_\star t) \\
{ z_\star}=0 \, .
\end{eqnarray}
I.e., in aligned systems, the planet re-encounters the same stellar wind property once every $2 \pi /(\Omega_{\rm orb}+\Omega_{\star})$, as long as the stellar magnetic field does not evolve significantly.

%%%%%%%%%%%%%%%%%%%%%%%%%%%%%%%%%%
\section{Interconnecting loop model}
A third star-planet scenario we also explore in this paper is based on the idea presented by \citet{2013A&A...557A..31L}, in which a magnetic loop connecting the star and the planet would suffer a rupture after being stretched by the planet. The idea is that one footpoint of the loop sits at the surface of the star, while the other footpoint lies at the surface of the planet.  As the planet moves through its orbit, the magnetic loop is stretched (stressed) and when it breaks, it can release energy, which travels towards the star, giving rise for example to anomalous hot spots at the stellar surface \citep{2005ApJ...622.1075S, 2019NatAs...3.1128C}. This is the scenario that \citet{2022MNRAS.512.4556S} named `stretch and break'; in their most recent work on star-planet interactions in the GJ\,436 system, this model was also investigated by \citet{2023AJ....165..146L}. In this scenario, the model assumes that the stellar closed corona extends up to the distance of the planet, which is not the case in the stellar wind models we presented here -- as showed in Section \ref{sec.spaceweather}, the stellar wind stretches open the magnetic field lines of the star at  distances within the planetary orbital distance. Additionally, the interconnecting loop scenario assumes that the stellar magnetic field is  close to a potential field, so that it can topologically reconnect with the magnetic field of the planet that is also potential. In the presence of a stellar wind though, the coronal magnetic field departs from a potential state \citep{2011MNRAS.412..351V}. 

\begin{figure}
    \centering
    \includegraphics[width=.48\textwidth]{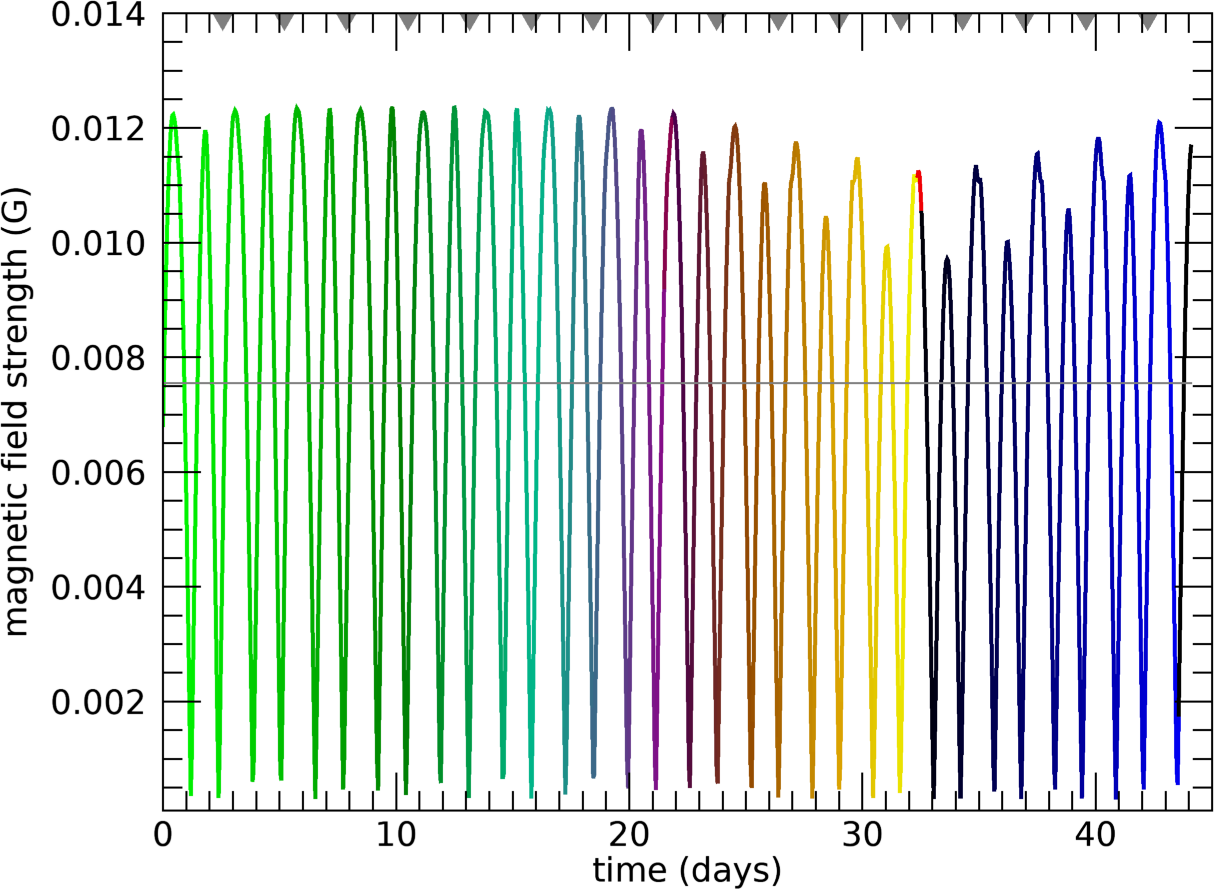} \\
    \includegraphics[width=.48\textwidth]{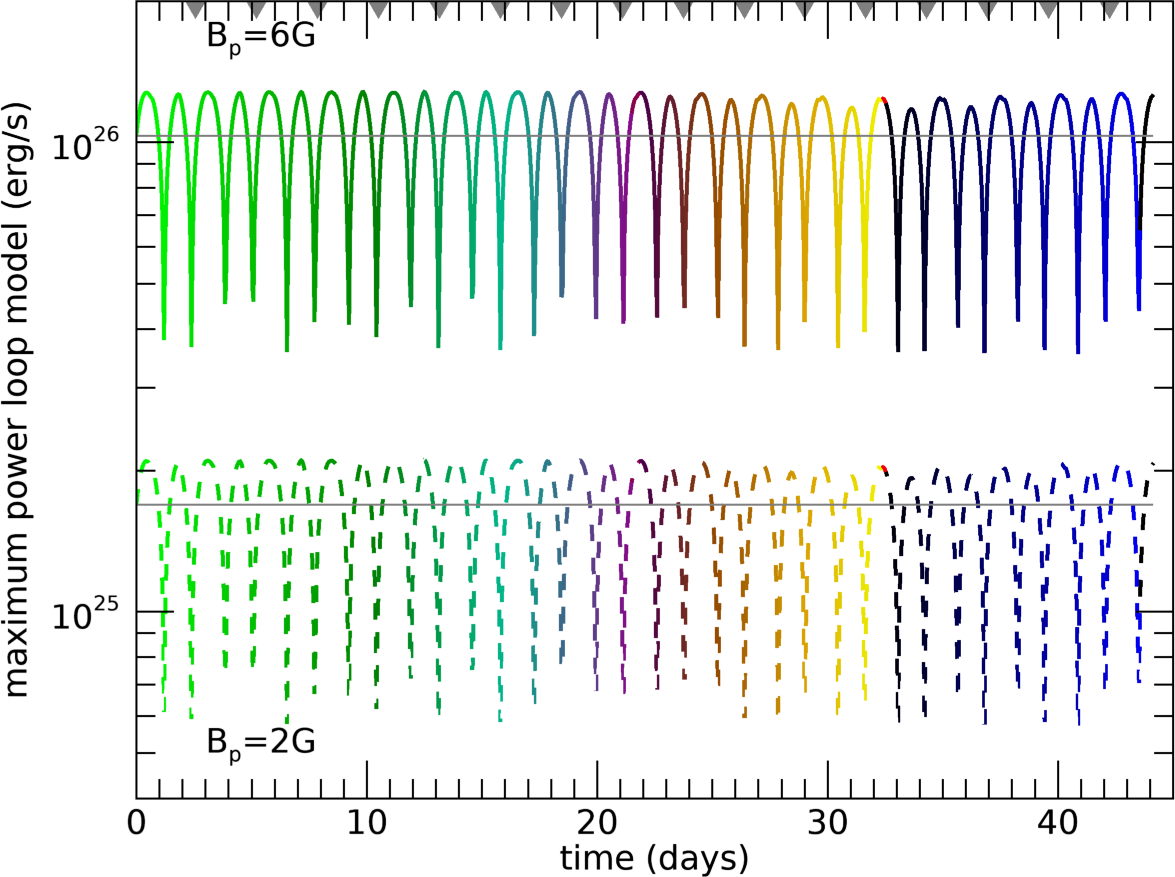}
    \caption{Top: magnetic field of the star at the orbital distance of the planet, assuming a potential field extrapolation, with a source surface of $\simeq 15 R_\star$ (in this case, no wind model is used). Bottom: the maximum dissipated power in the interconnecting loop model, assuming a polar planetary magnetic field strength (dipole) of 2~G (dashed) and 6~G (solid). The colour scheme follows the colour bar shown in Figure \ref{fig.pl_trajectory}.  {The grey arrows in the upper horizontal axes indicate approximate times of mid-transits.}}
    \label{fig.loop.model}%
\end{figure}

To explore the interconnecting loop scenario, we run a model without the presence of a stellar wind, in which we assume  that the stellar magnetic field remains potential and closed up to the orbit of the planet. To compute the field strength at the orbit of the planet, we use a potential field source surface extrapolation. In this model, the source surface represents the distance above which the magnetic field lines become open. To enforce that the stellar magnetic field lines are closed up to the planet's orbit, we choose a source surface of $\simeq 15 R_\star$, i.e., beyond the planet's orbit. Using the surface magnetic map from Figure \ref{fig.zdi} and our potential field model,  the magnetic field of the star at the orbit of the planet is shown in the top panel of Figure  \ref{fig.loop.model}, where we see that the coronal magnetic field is on average $7.5\times 10^{-3}$~G.

%int, Bp 1: ave, min, max:    0.00755095  0.000290509    0.0123858

With this, we can then compute the power released in the interconnecting loop model using \citep{2013A&A...557A..31L}
 \begin{equation}\label{eq.nuccio2}
 \mathcal{P}_{\rm sb} = (2 \pi R_p^2) f_{\rm open}  \left[ \frac{B_p^2 v_K}{4\pi} \right] \, , 
 \end{equation}
 where the Poynting flux (term within brackets) is now across the planet. The fraction of the planetary surface area that has open magnetic field lines  \citep{2011ApJ...730...27A} and thus have interconnecting magnetic loops is
 \begin{equation}
 f_{\rm open} = 1 - \left(1-\frac{3\zeta^{1/3}}{2+\zeta}\right)^{1/2} \, 
 \end{equation} 
with $\zeta = B/B_p$. Here, $B$ is the stellar magnetic field extrapolated out to the orbit of GJ\,436\,b and $B_p$, as before, is the assumed polar magnetic field strength of a dipolar planetary magnetic field. In the interconnecting loop scenario, the atmosphere of the planet is assumed to be ionised down to the surface, where the planetary magnetic field is $B_p$. If the atmosphere becomes neutral at a certain height above the surface, then the value of $B_p$ is that at such a height, which is smaller than the surface field. Therefore, Equation (\ref{eq.nuccio2}) provides an upper limit of the power released in the interaction. Assuming $B_p=2$~G, we find that the power released in the interconnecting loop model is on average $1.7\times 10^{25}$~erg~s$^{-1}$. For a magnetic field of $B_p = 6$~G, the power released increases to $10\times 10^{25}$~erg~s$^{-1}$ (this value is similar to the maximum power derived by \citealt{2023AJ....165..146L}). This is shown in the bottom panel of Figure \ref{fig.loop.model}.

We confirm the findings of previous studies \citep{2019NatAs...3.1128C, 2022MNRAS.512.4556S}, which showed that the interconnecting loop scenario provides the largest maximum power among the three scenarios explored in this paper (see the first two scenarios in Section \ref{sec.spi}), and is also in line with the estimates provided in \citet{2013A&A...557A..31L}. According to the interconnecting loop scenario, a planetary magnetic field  $B_p \lesssim 6$~G can reproduce the maximum power observed in star-planet interactions in the GJ\,436 system \citep{2023AJ....165..146L}. 

%%%% End of aa.dem
\end{document}